\documentstyle[aps]{revtex}

\begin{document}

\title{Translation-covariant Markovian master equation for a test
  particle in a quantum fluid}
\author{Bassano~Vacchini\footnote{E-mail: bassano.vacchini@mi.infn.it}}
\address{Dipartimento di Fisica  
dell'Universit\`a di Milano and Istituto Nazionale di Fisica  
Nucleare, Sezione di Milano, Via Celoria 16, I-20133, Milan,  
Italy}  
\date{\today}
\maketitle
\begin{abstract}
A recently proposed master equation in the Lindblad
form is studied with respect to covariance properties and existence 
of a stationary solution. The master equation describes the
interaction of a test particle with a quantum fluid, the so-called
Rayleigh gas, and is characterized by the appearance of a two-point
correlation function known as dynamic structure factor, which
reflects symmetry and statistical mechanics properties of the
fluid. In the case of a free gas all relevant physical
parameters, such as fugacity, ratio between the masses, momentum
transfer and energy transfer are put into evidence, giving an exact
expansion of the dynamic structure factor. The limit in which 
these quantities are small is then considered. In particular in the
Brownian limit a Fokker-Planck equation is obtained in which the
corrections due to quantum statistics can be explicitly evaluated and
are given in terms of the Bose function $g_0 (z)$ and the Fermi
function $f_0 (z)$.
\end{abstract}
\section{INTRODUCTION} 
\label{uno} 
The study of dissipative systems in a quantum mechanical framework is
a subject of major  interest for many physical communities 
especially in connection with
applications. Recently however the subject has gained new interest
also for physicists concerned with foundations of quantum mechanics,
due to the relevance of the notion of decoherence as a gateway between 
classical and quantum world~\cite{Kiefer}. This interest is strongly
supported and partially motivated by a spectacular improvement in many 
experimental techniques useful for handling with great precision
single- or few-particle systems, checking for coherence properties in
their dynamical evolution~\cite{Bonifacio}. In this connection 
models for quantum dissipation determined by the symmetry properties
of the microphysical interaction and by symmetry and statistical
mechanics properties of the environment could be of interest for a
large class of phenomena. In the Markovian limit quantum-dynamical
semigroups~\cite{Alicki} seem the most natural quantum mechanical
framework for the description of dissipative systems~\cite{Spohn} and
a lot of work has been done in this direction, both at rigorous and
formal level, especially with reference to the structural result of
Lindblad, which fixes the form of the generator of a completely
positive quantum-dynamical semigroup in the case of norm
continuity~\cite{Lindblad}. In this paper we will study in detail some 
structural properties of a recently proposed Markovian
master equation for the description of the dynamics of a test particle 
in a fluid~\cite{art1-art2,art3,art4,reply}, the so-called Rayleigh
gas~\cite{Spohn}. This simple but realistic model is of particular
interest in statistical mechanics, being a paradigmatic example which
opens the way to the study of interacting truly many-body systems. In
the quantum regime a feature of independent interest is the relevance
of quantum statistics of particles making up the fluid, especially in
connection with the recent experimental realization of almost
non-interacting degenerate gas samples of both Bose and Fermi
particles~\cite{bec-fermi}. The considered master equation was derived 
by assuming a translationally invariant interaction between the test
particle and a homogeneous fluid, and has the general structure of
generator of translation-covariant dynamical semigroups considered
in~\cite{HolevoTI}. In this way a direct physical interpretation
arises, at least in a particular case, of the general structure given
in~\cite{HolevoTI}. Further symmetry 
or equilibrium properties, which are of fundamental relevance in order 
to determine the realm of validity of a given master
equation~\cite{reply}, are shown to be a direct consequence of
particular physical features of the environment, embodied in a
specific two-point correlation function, the so-called
dynamic structure factor~\cite{Lovesey}. 
\par
The paper is organized as
follows: in Section~\ref{due} the master equation is introduced and the
property of covariance with respect to translations and rotations is
considered with reference to the corresponding symmetries of the
environment embodied in its dynamic structure factor, moreover the
existence of a stationary solution is proved, provided the environment 
is in a $\beta$-KMS state~\cite{KMS}; in Section~\ref{tre} the case of a free 
quantum gas is considered, the dynamic structure factor is explicitly
calculated, its exact expansion with respect to the relevant physical
parameters is obtained and in the limit of small fugacity $z$ the
expression for Maxwell-Boltzmann particles is recovered; in
Section~\ref{quattro} 
the Brownian limit in which the test particle is much heavier than the 
particles making up the gas is dealt with, together with the limit of
small momentum transfer, leading from the master equation to a
Fokker-Planck equation, strongly dependent on the statistics of the
gas; in Section~\ref{cinque} 
the obtained results are briefly summarized and discussed.
\section{GENERAL FEATURES OF THE MASTER EQUATION}  
\label{due} 
Let us recall the general expression of the master equation proposed
in~\cite{art3,art4} for the description of the motion of a test
particle in a homogeneous fluid supposed to be at equilibrium, whose 
properties we are going to study. The obtained result is based on a
scattering theory derivation, assuming a translationally invariant
interaction in terms of two-particle collisions, and is expected to be 
valid on a time scale much longer than the relaxation time of the
environment. In the Schr\"odinger picture the master equation is given by
\begin{equation}
  \label{me}
        {  
        d {\hat \varrho}  
        \over  
                      dt
        }  
        =
        {\cal M} [\hat \varrho]
        =
        -
        {i \over \hbar}
        [{\hat H}_0
        ,
        {\hat \varrho}
        ]
        +
        {\cal L} [\hat \varrho]
\end{equation}
where ${\hat H}_0={\hat {\bbox{p}}}^2 / 2M$ is the Hamiltonian
of the free particle, $M$ being the mass of the test particle, while
the dissipative part is given by the following mapping with a Lindblad 
structure
\begin{equation}
  \label{l}
        {\cal L}[\cdot]=
        \int_{{\bf R}^3}  d\mu ({\bbox{q}}) \,  
        \Biggl[
        \hat{U} ({\bbox{q}})
        \sqrt{
        S({\bbox{q}},{\hat {\bbox{p}}})
        }
        \cdot
        \sqrt{
        S({\bbox{q}},{\hat {\bbox{p}}})
        }
        \hat{U}^{\dagger} ({\bbox{q}})
        -
        \frac 12
        {
        \left \{
        S({\bbox{q}},{\hat {\bbox{p}}}),
        \cdot
        \right \}
        }
        \Biggr],  
\end{equation}
where the integral is over the parameter space of the translation
group in momentum space in three dimensions, the parameter
${\bbox{q}}$ being the momentum transferred in a collision,
$        
 \{
        \hat{A},\hat{B}
 \}
$ denotes the Jordan product $\hat{A} \circ
\hat{B}=\hat{A}\hat{B}+\hat{B}\hat{A}$, and the operator ${\hat
  {\bbox{p}}}$ is the  
generator of translations. The unitary operators 
$\hat{U} ({\bbox{q}})=e^{{i\over\hbar}{\bbox{q}}\cdot{\hat
    {\bbox{x}}}}$, ${\bbox{q}} \in {\bf R}^3$, are the generators of
translations in momentum space or boosts. The positive measure $d\mu
({\bbox{q}})$ is given by
\begin{equation}
  \label{misura}
  d\mu ({\bbox{q}})=        {2\pi \over\hbar}
        (2\pi\hbar)^3
        n
        {
        | \tilde{t} (q) |^2
        }
        d^3\!
        {\bbox{q}}
        ,
\end{equation}
thus being invariant under both rotations and translations. In fact
$n$ is the particle density in the macroscopic system, and the
function $\tilde{t} (q)$, given by
\begin{displaymath}
  \tilde{t} (q)=
        \int_{{\bf R}^3} 
        d^3\!
        {\bbox{x}}
        \,
        {
        e^{{i\over\hbar}{\bbox{q}}\cdot{{\bbox{x}}}}  
        \over  
        (2\pi\hbar)^3  
        }
        t (x)
        ,
\end{displaymath}
where $q=|{\bbox{q}}|$ and $x=|{\bbox{x}}|$,
is the Fourier transform with respect to the
transferred momentum ${\bbox{q}}$ of the T matrix describing the
translationally and rotationally invariant interaction between the
test particle and the particles of the fluid, which is supposed to be
energy independent.
The function $S({\bbox{q}},{\bbox{p}})$, which
appears operator-valued in (\ref{l}), is a positive two-point
correlation 
function known as dynamic structure factor~\cite{Lovesey}, given by
the Fourier transform with respect to energy transfer $E
({\bbox{q}},{\bbox{p}})$ and momentum transfer ${\bbox{q}}$ of
the time dependent pair correlation function of the fluid
\begin{equation}
  \label{dsf}
  S({\bbox{q}},{\bbox{p}}) = 
        {  
        1  
        \over  
         2\pi\hbar
        }  
        \int_{{\bf R}} dt 
        {\int_{{\bf R}^3} d^3 \! {\bbox{x}} \,}        
        e^{
        {
        i
        \over
         \hbar
        }
        [E ({\bbox{q}},{\bbox{p}}) t -
        {\bbox{q}}\cdot{\bbox{x}}]
        } 
        G ({\bbox{x}},t)
        ,
\end{equation}
where $G ({\bbox{x}},t)$ is the time dependent pair correlation
function
\begin{equation}
  \label{g}
  G ({\bbox{x}},t)=\frac{1}{N}
        {\int_{{\bf R}^3} d^3 \! {\bbox{y}} \,}
        \left \langle  
         N({\bbox{y}})  
         N({\bbox{y}}+{\bbox{x}},t)  
        \right \rangle, 
\end{equation}
$
        N({\bbox{y}})
$
being the operator density of particles in the fluid and 
$
        \left \langle  
\ldots
        \right \rangle
$
the ensemble average with respect to the state of the macroscopic
system. The expression $E ({\bbox{q}},{\bbox{p}})$ gives the energy
transfer in a collision where the test particle changes its momentum
from ${\bbox{p}}$ to ${\bbox{p}}+{\bbox{q}}$, so that
\begin{equation}
  \label{E}
  E ({\bbox{q}},{\bbox{p}})=E_{p+q} - E_{p}
=
{
q^2
\over
   2M
}
+
{
        {\bbox{p}}
        \cdot
        {\bbox{q}}
\over
M
}
.
\end{equation}
Note that in (\ref{dsf}) we have used as variables momentum and energy
transferred to the test particle. In the sequel we will use both the
equivalent notations $S ({\bbox{q}},{\bbox{p}})$ and $S ({\bbox{q}},E
({\bbox{q}},{\bbox{p}}))\equiv  S ({\bbox{q}},E)$, according to
convenience. The dynamic structure factor is a very important
physical quantity, giving the spectrum of spontaneous fluctuations of
the system, and it is of direct experimental access: in fact, as first
obtained by van Hove in a fundamental work~\cite{vanHove}, it is directly 
related to the energy dependent differential cross-section per target particle describing
scattering of a microscopic probe off a macroscopic sample through the formula
\begin{equation}
  \label{diff}
  \frac{d^2 \sigma}{d\Omega_{p'} dE_{p'}}  =
\left({2\pi\hbar}\right)^6
\left(\frac{M}{2\pi\hbar^2}\right)^2
\frac{p'}{p}
        {
        | \tilde{t} (q) |^2
        }
  S ({\bbox{q}},E)
,  
\end{equation}
referring to scattering of the microscopic probe from ${\bbox{p}}$ to
${\bbox{p}}' = {\bbox{p}}+{\bbox{q}}$. The dynamic
structure factor 
is given in (\ref{diff}) as a function of momentum and energy
transfer, which are the measured quantities in scattering experiments, 
the energy $E$ being related to ${\bbox{q}}$ and ${\bbox{p}}$ through
(\ref{E}). The appearance of the dynamic structure factor in
(\ref{diff}) gives the physical reason for its being positive
definite for every system. 
The main point in (\ref{me}) and (\ref{l}) is the determination of the
specific expressions for the measure $d\mu ({\bbox{q}})$ given by
(\ref{misura}), and of the operator valued function $S
({\bbox{q}},\cdot)$, given by (\ref{dsf}), which can only be
obtained on the basis of a microphysical derivation of the equation,
relying on some physical model. 
A general structure encompassing
(\ref{l}) has already been considered by Holevo in a purely mathematical
context, studying the general structure of generators of 
translation-covariant dynamical semigroups~\cite{HolevoTI}. In
particular Holevo has proven that when the generator is bounded it
must have a structure of the form (\ref{me})
with ${\cal L}$ given by (\ref{l}) provided all operators appearing in
(\ref{me}) and (\ref{l}) are bounded~\cite{HolevoTI} (see
also~\cite{ManitaHPA,Manita}, 
where further restrictions to the structure of (\ref{l}) appear, to be
discussed later on), and allowing, instead of the structure
\begin{displaymath}
        \hat{U} ({\bbox{q}})
        \sqrt{
        S({\bbox{q}},{\hat {\bbox{p}}})
        }
        \cdot
        \sqrt{
        S({\bbox{q}},{\hat {\bbox{p}}})
        }
        \hat{U}^{\dagger} ({\bbox{q}})
\end{displaymath}
appearing in (\ref{l}), where $S({\bbox{q}},{\hat {\bbox{p}}})$ is
self-adjoint and positive, the more 
general structure
\begin{displaymath}
        \hat{U} ({\bbox{q}})
        {
        L({\bbox{q}},{\hat {\bbox{p}}})
        }
        \cdot
        {
        L^{\dagger}({\bbox{q}},{\hat {\bbox{p}}})
        }
        \hat{U}^{\dagger} ({\bbox{q}})
.  
\end{displaymath}
If the generator is unbounded also diffusion terms of the form
considered in  (\ref{qd}) may appear, and the operators appearing in
(\ref{me}) and (\ref{l}) may be unbounded (see~\cite{HolevoTI} for further details).
In the general model considered here the Hamiltonian is given by the
unbounded operator
${\hat H}_0={\hat {{\bbox{p}}}}^2 / 2M$, while 
the remaining part of the generator is determined by the
explicit expression of the physical quantities $\tilde{t} (q)$ and $S({\bbox{q}},{\bbox{p}})$,
depending on the specific model for the fluid.
\par
We now consider the
behavior of (\ref{me}) with respect to symmetry transformations. Let us 
consider a locally-compact group $G$ and a unitary representation 
$\hat{U} (g)$, with
$g \in G$, on the Hilbert space of the
system. Following~\cite{HolevoTI} we say that a mapping 
${\cal M}$ in the Schr\"odinger picture is $G$-covariant if it
commutes with the mapping
${\cal U}_g [\cdot]=\hat{U} (g) \cdot \hat{U}^{\dagger} (g)$ 
for all $g \in G$
\begin{equation}
  \label{um}
  {\cal M}[{\cal U}_g[\cdot]]={\cal U}_g[{\cal M}[\cdot]] 
  .
\end{equation}
Let us first consider the case of spatial translations. Then the
unitary operators are given by
$
\hat{U} ({\bbox{a}})=e^{-{i\over\hbar}{\bbox{a}}\cdot{\hat
    {\bbox{p}}}}
$
with ${\bbox{a}} \in {{\bf R}^3}$ and exploiting
\begin{displaymath}
[\hat{U} ({\bbox{a}}),{\hat H}_0]  =0, \qquad 
[\hat{U} ({\bbox{a}}),S({\bbox{q}},{\hat {\bbox{p}}})]  =0 
,
\end{displaymath}
together with the Weyl CCR
\begin{displaymath}
\hat{U} ({\bbox{q}})\hat{U} ({\bbox{a}})=\hat{U} ({\bbox{a}})\hat{U}
({\bbox{q}}) e^{{i\over\hbar}{\bbox{a}}\cdot{    {\bbox{q}}}} 
,
\end{displaymath}
one immediately has that the mapping ${\cal M}$ given by (\ref{me}) is
translation-covariant. This property goes back to homogeneity of the
fluid and translational invariance of the interaction, as can be seen
in the derivation of the master equation~\cite{art3,art4}. We then
consider invariance under rotations, so that the relevant set of
unitary operators takes the form
$\hat{U} (R)$, with $R\in
{\bf SO} (3)$. 
In this case exploiting
$[\hat{U} (R),{\hat H}_0]  =0$ and the relations
\begin{displaymath}
  \hat{U}^{\dagger} (R)\hat{U} ({\bbox{q}})\hat{U} (R)=\hat{U}
  (R^{-1}{\bbox{q}}), \qquad
  \hat{U}^{\dagger} (R)\hat{\bbox{p}}\hat{U} (R)=R\hat{\bbox{p}},
\end{displaymath}
one has that ${\cal M}$ is rotation-covariant provided the
dynamic structure factor satisfies for $R \in {\bf SO} (3)$
\begin{equation}
  \label{ruota}
  S(R{\bbox{q}},R{ {\bbox{p}}})=S({\bbox{q}},{ {\bbox{p}}})
.
\end{equation}
In fact if (\ref{ruota}) holds one has
\begin{eqnarray*}
        {\cal M}[{\cal U}_R [\cdot]] 
        &=&
        \int_{{\bf R}^3}  d\mu ({\bbox{q}}) \,  
        \hat{U} (R)\Biggl[
        \hat{U} (R^{-1}{\bbox{q}})
        \sqrt{
        S({\bbox{q}},R{\hat {\bbox{p}}})
        }
        \cdot
        \sqrt{
        S({\bbox{q}},R{\hat {\bbox{p}}})
        }
        \hat{U}^{\dagger} (R^{-1}{\bbox{q}})
        -
        \frac 12 
        \left \{
        S({\bbox{q}},R{\hat {\bbox{p}}}),
        \cdot
        \right \}
        \Biggr] \hat{U}^{\dagger} (R)
        \\
        &=&
        \int_{{\bf R}^3}  d\mu ({\bbox{q}}) \,  
        \hat{U} (R) \Biggl[
        \hat{U} ({\bbox{q}})
        \sqrt{
        S({\bbox{q}},{\hat {\bbox{p}}})
        }
        \cdot
        \sqrt{
        S({\bbox{q}},{\hat {\bbox{p}}})
        }
        \hat{U}^{\dagger} ({\bbox{q}})
        -
        \frac 12
        \left \{
        S({\bbox{q}},{\hat {\bbox{p}}}),
        \cdot
        \right \}
        \Biggr]\hat{U}^{\dagger} (R)
        \\
        &=&
        {\cal U}_R[ {\cal M}[\cdot]] 
        .
\end{eqnarray*}
On the other hand (\ref{ruota}) is directly linked to rotational invariance
of the surrounding environment, as one can see observing that
$E({\bbox{q}},{{\bbox{p}}})$ as given by (\ref{E}) satisfies 
\begin{displaymath}
  E(R{\bbox{q}},R{ {\bbox{p}}})=E({\bbox{q}},{ {\bbox{p}}})
  ,
\end{displaymath}
and consider the identity
\begin{eqnarray*}
S(R{\bbox{q}},R{ {\bbox{p}}}) &=&
        {  
        1  
        \over  
         2\pi\hbar
        }  
        \int_{{\bf R}} dt 
        {\int_{{\bf R}^3} d^3 \! {\bbox{x}} \,}        
        e^{
        {
        i
        \over
         \hbar
        }
        [E (R{\bbox{q}},R{\bbox{p}}) t -
        R{\bbox{q}}\cdot{\bbox{x}}]
        } 
        G ({\bbox{x}},t)
        \\
        &=&
        {  
        1  
        \over  
         2\pi\hbar
        }  
        \int_{{\bf R}} dt 
        {\int_{{\bf R}^3} d^3 \! {\bbox{x}} \,}        
        e^{
        {
        i
        \over
         \hbar
        }
        [E ({\bbox{q}},{\bbox{p}}) t -
        {\bbox{q}}\cdot{\bbox{x}}]
        } 
        G (R{\bbox{x}},t)
\end{eqnarray*}
so that (\ref{ruota}) holds if and only if the pair correlation
function is invariant under rotations,
\begin{displaymath}
       G (R{\bbox{x}},t)=G ({\bbox{x}},t).
\end{displaymath}
\par
In order to proceed further and consider the
existence of stationary solutions we make the natural assumption that
the state of the macroscopic system, with respect to which the expectation value
in (\ref{g}) is calculated, is a $\beta$-KMS state, so that the
relation
\begin{equation}
  \label{beta}
  \langle \hat{A} (w)\hat{B}\rangle=\langle\hat{B} \hat{A} (w+i\hbar
  \beta)\rangle 
\end{equation}
holds.
This in turn implies that the dynamic structure factor $S
({\bbox{q}},{\bbox{p}})$ satisfies an identity known as {\it detailed
  balance condition\/}~\cite{Lovesey}
\begin{equation}
  \label{dbcp}
  S
({\bbox{q}},{\bbox{p}})
=
e^{-\beta ({
q^2
\over
   2M
}
+
{
        {\bbox{p}}
        \cdot
        {\bbox{q}}
\over
M
}
)}
S
(-{\bbox{q}},{\bbox{p}}+{\bbox{q}}),
\end{equation}
usually expressed in terms of the dependence on the transferred energy
\begin{equation}
  \label{dbce}
  S
({\bbox{q}},E
({\bbox{q}},{\bbox{p}}))
=
e^{-\beta E
({\bbox{q}},{\bbox{p}})
}
S
(-{\bbox{q}},-E
({\bbox{q}},{\bbox{p}})),
\end{equation}
the sign of the exponential being determined by the fact that we are
considering momentum and energy transferred to the particle. It will
prove useful for further considerations to introduce a symmetrized
version of the dynamic structure factor, given by
\begin{equation}
  \label{dsfsimme}
  \tilde{S}
({\bbox{q}},E)=e^{\frac{\beta}{2}E}S
({\bbox{q}},E)
\end{equation}
or equivalently 
\begin{equation}
  \label{dsfsimmp}
  \tilde{S}
({\bbox{q}},{\bbox{p}})=e^{\frac{\beta}{2}
\left(
{
q^2
\over
   2M
}
+
{
        {\bbox{p}}
        \cdot
        {\bbox{q}}
\over
M
}
\right)}S
({\bbox{q}},{\bbox{p}}),
\end{equation}
satisfying instead of (\ref{dbce}) the more symmetric
\begin{displaymath}
  \tilde{S}
({\bbox{q}},E)=\tilde{S}
(-{\bbox{q}},-E),
\end{displaymath}
so that (\ref{dbcp}) becomes
\begin{equation}
  \label{5}
  \tilde{S}({\bbox{q}},{\bbox{p}})=
  \tilde{S}(-{\bbox{q}},{\bbox{p}}+{\bbox{q}})
  .
\end{equation}
We now look for a stationary solution of (\ref{me}), given that the
environment is in a $\beta$-KMS state, so that due to (\ref{beta}) the 
dynamic structure factor satisfies the detailed balance condition as 
in (\ref{dbcp}). According to translation-covariance of the generators 
we look for a solution invariant under translation, of the
form $\rho (\hat{\bbox{p}})$. Since
$[\hat{H}_0,\rho (\hat{\bbox{p}})]=0$, $\rho (\hat{\bbox{p}})$ will be 
a stationary solution of (\ref{me}) provided
\begin{eqnarray}
  \label{1}
        {\cal L}[\rho (\hat{\bbox{p}})]
        &=&
        \int_{{\bf R}^3}  d\mu ({\bbox{q}}) \,  
        \Biggl[
        \hat{U} ({\bbox{q}})
        {
        S({\bbox{q}},{\hat {\bbox{p}}})
        }
        \rho (\hat{\bbox{p}})
        \hat{U}^{\dagger} ({\bbox{q}})
        -
        S({\bbox{q}},{\hat {\bbox{p}}})
        \rho (\hat{\bbox{p}})
        \Biggr]
        \\
        &=&
        \int_{{\bf R}^3}  d\mu ({\bbox{q}}) \,  
        \Biggl[
        S({\bbox{q}},{\hat {\bbox{p}}}-{\bbox{q}})
        \rho (\hat{\bbox{p}}-{\bbox{q}})
        -
        S({\bbox{q}},{\hat {\bbox{p}}})
        \rho (\hat{\bbox{p}})
        \Biggr]
        =0
        .
        \nonumber
\end{eqnarray}
Introducing the function
\begin{equation}
  \label{a}
  A ({\bbox{q}},{\bbox{p}})=
        S({\bbox{q}},{ {\bbox{p}}}-{\bbox{q}})
        \rho ({\bbox{p}}-{\bbox{q}})
        -
        S({\bbox{q}},{ {\bbox{p}}})
        \rho ({\bbox{p}})
\end{equation}
the requirement (\ref{1}) becomes
\begin{equation}
  \label{2}
  \int_{{\bf R}^3}  d\mu ({\bbox{q}}) \,  
  A ({\bbox{q}},{\bbox{p}})=0.
\end{equation}
A sufficient condition for (\ref{2}) to be valid is that $A
({\bbox{q}},{\bbox{p}})$ be an odd function in ${\bbox{q}}$, and we
shall see that this is exactly the case if $\rho (\hat{\bbox{p}})$ has 
the canonical structure $\rho_0 (\hat{\bbox{p}})=e^{-\beta {
{\hat {\bbox{p}}}^2
\over
     2M
}
}$, with $M$ the mass of the test particle and $\beta$ the inverse
temperature of the macroscopic system, as it is to be expected on
physical grounds. 
Let $\rho_0 (\hat{\bbox{p}})=e^{-\beta {
{\hat {\bbox{p}}}^2
\over
     2M
}
}$, then
\begin{displaymath}
  \rho_0 ({\bbox{p}}-{\bbox{q}})=\rho_0 ({\bbox{p}})
  e^{-\beta ({
      q^2
      \over
      2M
      }
    -
    {
      {\bbox{p}}
      \cdot
      {\bbox{q}}
      \over
      M
      }
    )}  
  ,
\end{displaymath}
so that
\begin{displaymath}
  A ({\bbox{q}},{\bbox{p}})
  =
  \rho_0 ({\bbox{p}})
  \left[
    S({\bbox{q}},{ {\bbox{p}}}-{\bbox{q}})
  e^{-\beta ({
      q^2
      \over
      2M
      }
    -
    {
      {\bbox{p}}
      \cdot
      {\bbox{q}}
      \over
      M
      }
    )} 
  -
  S({\bbox{q}},{ {\bbox{p}}})
  \right]
\end{displaymath}
and exploiting (\ref{dbcp})
\begin{displaymath}
  A ({\bbox{q}},{\bbox{p}})
  =
  \rho_0 ({\bbox{p}})
  \left[
  S(-{\bbox{q}},{{\bbox{p}}})
  -
  S({\bbox{q}},{ {\bbox{p}}})
  \right]
,  
\end{displaymath}
which is manifestly odd in ${\bbox{q}}$.
\par
As mentioned above the structure given in (\ref{l}) is a particular 
realization of the general expression considered
in~\cite{HolevoTI}, however it does not meet the more stringent
requirements exhibited by the dissipative mapping considered
in~\cite{Manita}. According to~\cite{HolevoTI} these
requirements are unnecessary if one is looking for the most general
translation-covariant generator and we shall show that in the present
framework they would lead to unphysical results. In fact the structure 
proposed in~\cite{Manita} for the dissipative part would take in the
Schr\"odinger picture the form
\begin{equation}
  \label{manita}
        {\cal L}[\cdot]=
        \int_{{\bf R}^3}  d\mu ({\bbox{q}}) \,  
        \Biggl[
        P({\bbox{q}},{\hat {\bbox{p}}})
        \cdot
        P^{\dagger}({\bbox{q}},{\hat {\bbox{p}}})
        -
        \frac 12
        \left \{
        P^{\dagger}({\bbox{q}},{\hat {\bbox{p}}})
        P({\bbox{q}},{\hat {\bbox{p}}}),
        \cdot
        \right \}
        \Biggr],  
\end{equation}
with the further requirement
\begin{equation}
  \label{cond}
  P^{\dagger}({\bbox{q}},{\hat {\bbox{p}}})=P(-{\bbox{q}},{\hat {\bbox{p}}}),
\end{equation}
which in our case, since
\begin{equation}
  \label{4}
  P({\bbox{q}},{\hat {\bbox{p}}})=
        \hat{U} ({\bbox{q}})
        \sqrt{
        S({\bbox{q}},{\hat {\bbox{p}}})
        }
\end{equation}
is not satisfied, being equivalent to the requirement
\begin{displaymath}
        \hat{U} ({\bbox{q}})
        \sqrt{
        S({\bbox{q}},{\hat {\bbox{p}}})
        }
      =
        \hat{U} ({\bbox{q}})
        \sqrt{
        S(-{\bbox{q}},{\hat {\bbox{p}}}+{\bbox{q}})
        }
      ,
\end{displaymath}
which does not hold due to the presence of the factor
$e^{-\beta ({
q^2
\over
   2M
}
+
{
        {\bbox{p}}
        \cdot
        {\bbox{q}}
\over
M
}
)}$ in (\ref{dbcp}).
If instead of the dynamic structure factor  $S({\bbox{q}},{\bbox{p}})$
one would consider the symmetrized dynamic structure factor
$\tilde{S}({\bbox{q}},{\bbox{p}})$ given by (\ref{dsfsimmp}), so that
$P({\bbox{q}},{\bbox{p}})$ in (\ref{4}) would be replaced by
\begin{displaymath}
  \tilde{P}({\bbox{q}},\hat{\bbox{p}})
  =
        \hat{U} ({\bbox{q}})
        \sqrt{
        \tilde{S}({\bbox{q}},{\hat {\bbox{p}}})
        }
  ,
\end{displaymath}
then according to (\ref{5}) the relation (\ref{cond}) would hold and
the dissipative mapping ${\cal L}$ would conform to the structure
proposed in~\cite{Manita}. In this case, however, one would have
the unphysical  
result that $\rho_0 (\hat{\bbox{p}})=e^{-\beta {
{\hat {\bbox{p}}}^2
\over
     2M
}
}$
is no more a stationary solution, because
\begin{displaymath}
    \tilde{A} ({\bbox{q}},{\bbox{p}})=
        \tilde{S}({\bbox{q}},{ {\bbox{p}}}-{\bbox{q}})
        \rho_0 ({\bbox{p}}-{\bbox{q}})
        -
        \tilde{S}({\bbox{q}},{ {\bbox{p}}})
        \rho_0 ({\bbox{p}})
\end{displaymath}
is no more odd in ${\bbox{q}}$. Other unphysical features linked to
the further restriction (\ref{cond}) will be considered in
Section~\ref{quattro}. Note that these features  
can only be discovered with reference to
a specific structure of translation-covariant generator determined by
starting from some microphysical model. In fact the restriction
(\ref{cond}), though unnecessary from a mathematical standpoint, could 
have proven as well interesting from a physical point of view, thus
suggesting the result of~\cite{Manita} as a useful starting point for
phenomenological approaches. This does not seem to be the case, in
fact the substitution ${S}({\bbox{q}},{ {\bbox{p}}})\rightarrow
\tilde{S}({\bbox{q}},{ {\bbox{p}}})$, natural in order to comply
with~\cite{Manita}, leads to unphysical results.
\section{EXACT EXPRESSION FOR A FREE QUANTUM GAS}  
\label{tre} 
It is of course of interest to analyze the mapping ${\cal M}$ given in 
(\ref{me}) for a model in which the dynamic structure factor of the
fluid can be explicitly calculated: this is the case for an ideal
quantum gas considered at finite temperature $T=1/\beta {\rm k}$,
where ${\rm k}$ is Boltzmann's constant, and obeying either Bose or
Fermi statistics. Apart from simplicity, the case of a free gas can be 
of interest also in view of the recent experimental realization of
dilute quantum samples of Bose or Fermi particles in the degenerate
regime~\cite{bec-fermi}. The dynamic structure factor for an ideal gas 
takes the form~\cite{Lovesey}
\begin{equation}
  \label{sbf}
        S_{\rm \scriptscriptstyle B/F}({\bbox{q}},{\bbox{p}})
        =
        \frac 1n
        \int_{{\bf R}^3} 
        {
        d^3 \!
        {\bbox{k}}
        \over
        (2\pi \hbar)^3
        }
        \,  
        \langle
        n_{k}
        \rangle_{\rm \scriptscriptstyle B/F}
        (1\pm
        \langle
        n_{k - q}
        \rangle_{\rm \scriptscriptstyle B/F}
        )
        \delta       
        \left(
        {({{\bbox{p}} + {\bbox{q}}})^2\over 2M}
        +
        {({{\bbox{k}} - {\bbox{q}}})^2\over 2m}
        -
        {{{\bbox{p}}}^2 \over 2M}
        -
        {{\bbox{k}}^2\over 2m}
        \right)
\end{equation}
where the indexes $B$ or $F$ and signs $+$ or $-$ refer to Bose or Fermi
statistics respectively, $M$ is the mass of the test particle, $m$ is the 
mass of the particles making up the gas, $n$ is the density of
particles in the gas and 
$
        \langle
        n_{k}
        \rangle_{\rm \scriptscriptstyle B/F}
$
is 
\begin{displaymath}
          \langle
        n_{k}
        \rangle_{\rm \scriptscriptstyle B/F}
        =
        {
        1
        \over  
        z^{-1}e^{\beta \epsilon_k}\mp 1
        }
      ,
      \qquad
      \epsilon_k = {{\bbox{k}}^2\over 2m}
\end{displaymath}
where $z$ is the fugacity of the gas, related to the chemical
potential $\mu$ by $z=e^{\beta\mu}$. For a Bose gas at finite
temperature $0\leq z<1$, while for a Fermi gas $z\geq 0$. The integral 
in (\ref{sbf}) can be explicitly calculated both for bosons and
fermions, giving the result (\ref{abf}) obtained in Appendix~\ref{aa}:
\begin{equation}
  \label{6}
        S_{\rm \scriptscriptstyle B/F}({\bbox{q}},{\bbox{p}})
        =
        \mp
        {
        1
        \over
         (2\pi\hbar)^3
        }
        {
        2\pi m^2
        \over
        n\beta q
        }
        {
        1
        \over
        1-
        e^{\beta E({\bbox{q}},{\bbox{p}})}
        }
        \log
        \left[
        {
        1\mp z
        \exp{
        \left[
        -{
        \beta
        \over
             8m
        }
        {
        (2mE({\bbox{q}},{\bbox{p}}) + q^2)^2
        \over
                  q^2
        }
        \right]
        }
        \over
        1\mp z
        \exp{
        \left[
        -{
        \beta
        \over
             8m
        }
        {
        (2mE({\bbox{q}},{\bbox{p}}) - q^2)^2
        \over
                  q^2
        }
        \right]
        }
        }
        \right],
\end{equation}
with $q=|{\bbox{q}}|$.
In the same way one can consider the case of a free gas of particles
satisfying Maxwell-Boltzmann statistics, thus having
\begin{equation}
  \label{smb}
        S_{\rm \scriptscriptstyle MB}({\bbox{q}},{\bbox{p}})
        =
        \frac 1n
        \int_{{\bf R}^3} 
        {
        d^3 \!
        {\bbox{k}}
        \over
        (2\pi \hbar)^3
        }
        \,  
        \langle
        n_{k}
        \rangle_{\rm \scriptscriptstyle MB}
        \delta       
        \left(
        {({{\bbox{p}} + {\bbox{q}}})^2\over 2M}
        +
        {({{\bbox{k}} - {\bbox{q}}})^2\over 2m}
        -
        {{{\bbox{p}}}^2 \over 2M}
        -
        {{\bbox{k}}^2\over 2m}
        \right)
        ,
\end{equation}
with
\begin{displaymath}
        \langle
        n_{k}
        \rangle_{\rm \scriptscriptstyle MB}
        =
        ze^{-\beta\epsilon_k},
\end{displaymath}
so that the integral in (\ref{smb}) can also be explicitly calculated
giving the expression (\ref{amb})
\begin{equation}
  \label{7}
        S_{\rm \scriptscriptstyle MB}({\bbox{q}},{\bbox{p}})
        =
        {
        1
        \over
         (2\pi\hbar)^3
        }
        {
        2\pi m^2
        \over
        n\beta q
        }
        z
        \exp\left[
        -{
        \beta
        \over
             8m
        }
        {
        (2mE({\bbox{q}},{\bbox{p}}) + q^2)^2
        \over
                  q^2
        }
        \right]
        .
\end{equation}
A convenient way to write (\ref{6}) and (\ref{7}) for later expansions 
is in terms of the function
\begin{equation}
  \label{77}
  \sigma({\bbox{q}},{\bbox{p}})=
\frac{1}{2q}
\left[
q^2+2\alpha M E({\bbox{q}},{\bbox{p}})
\right]
\end{equation}
where the ratio $\alpha=m/M$
between the masses of the particles of the gas and of the test
particle has been put into evidence, thus obtaining respectively
\begin{equation}
  \label{8}
        S_{\rm \scriptscriptstyle B/F}({\bbox{q}},{\bbox{p}})
        =
        \mp
        {
        1
        \over
         (2\pi\hbar)^3
        }
        {
        2\pi m^2
        \over
        n\beta q
        }
        {
        1
        \over
        1-
        \exp{
        \left[  
        {
        \beta
        \over
             2m
        }
        \left(
        2\sigma({\bbox{q}},{\bbox{p}})q -q^2
        \right)
        \right]
        }
        }
        \log
        \left[
        {
        1\mp z
        \exp{
        \left[  
        -{
        \beta
        \over
             2m
        }
        \sigma^2({\bbox{q}},{\bbox{p}})
        \right]
        }
        \over
        1\mp z
        \exp{
        \left[  
        -{
        \beta
        \over
             2m
        }
        (\sigma({\bbox{q}},{\bbox{p}}) - q)^2
        \right]
        }
        }
        \right]  
\end{equation}
and
        \begin{equation}
        \label{9}
        S_{\rm \scriptscriptstyle MB}({\bbox{q}},{\bbox{p}})
        =
        {
        1
        \over
         (2\pi\hbar)^3
        }
        {
        2\pi m^2
        \over
        n\beta q
        }
        z
        \exp{
        \left[  
        -{
        \beta
        \over
             2m
        }
        \sigma^2({\bbox{q}},{\bbox{p}})
        \right]
        }.
        \end{equation}
We have thus put into evidence all the physical parameters
which are of interest in
specifying the physical model under consideration and its range of
validity: ${\bbox{q}}$, $E$, $\alpha$ and $z$. In this perspective the
expression for a gas of 
Maxwell-Boltzmann particles can also be obtained as expected from the
dynamic structure factor for a Bose or Fermi gas in the limit of small 
fugacity $z$. In fact starting from (\ref{8}) and expanding the
logarithm up to first order in $z$ one has
\begin{eqnarray*}
        S_{\rm \scriptscriptstyle B/F}({\bbox{q}},{\bbox{p}},z\ll 1)
        &=&
        {
        1
        \over
         (2\pi\hbar)^3
        }
        {
        2\pi m^2
        \over
        n\beta q
        }
        {
        z
        \over
        1-
        \exp{
        \left[  
        {
        \beta
        \over
             2m
        }
        \left(
        2\sigma({\bbox{q}},{\bbox{p}})q -q^2
        \right)
        \right]
        }
        }
      \\
      &&
      \hphantom{        {
        1
        \over
         (2\pi\hbar)^3
        }
        {
        2\pi m^2
        \over
        n\beta q
        }
      }
    \times
        \left\{
        \exp{
        \left[  
        -{
        \beta
        \over
             2m
        }
        \sigma^2({\bbox{q}},{\bbox{p}})
        \right]
        }
      -
        \exp{
        \left[  
        -{
        \beta
        \over
             2m
        }
        (\sigma({\bbox{q}},{\bbox{p}}) - q)^2
        \right]
        }
        \right\}  
        \\
        &=&
        {
        1
        \over
         (2\pi\hbar)^3
        }
        {
        2\pi m^2
        \over
        n\beta q
        }
      z
        \exp{
        \left[  
        -{
        \beta
        \over
             2m
        }
        \sigma^2({\bbox{q}},{\bbox{p}})
        \right]
        }
        =
        S_{\rm \scriptscriptstyle MB}({\bbox{q}},{\bbox{p}})
        .
\end{eqnarray*}
Both (\ref{6}) and (\ref{7})or equivalently (\ref{8}) and (\ref{9})
are invariant under rotation, as one can see from the fact that they
depend on ${\bbox{q}}$ and ${\bbox{p}}$ only through
$E({\bbox{q}},{\bbox{p}})$ and the modulus $q$ of ${\bbox{q}}$, so
that
\begin{displaymath}
  S_{\rm \scriptscriptstyle B/F}(R{\bbox{q}},R{\bbox{p}})=S_{\rm
    \scriptscriptstyle B/F}({\bbox{q}},{\bbox{p}}) ,
  \qquad
  S_{\rm \scriptscriptstyle MB}(R{\bbox{q}},R{\bbox{p}})=S_{\rm
    \scriptscriptstyle MB}({\bbox{q}},{\bbox{p}}) ,
\end{displaymath}
thus leading to a rotation-covariant mapping ${\cal M}$ when
substituted in (\ref{me}). In order to grant the existence of the
stationary solution $\rho_0 (\hat{\bbox{p}})$ we have to check that
the obtained expressions satisfy the principle of detailed
balance. Starting from (\ref{6}) we have, setting for simplicity
$E({\bbox{q}},{\bbox{p}})=E$ and inverting the argument of the
logarithm
\begin{eqnarray*}
  e^{-\beta E}S_{\rm \scriptscriptstyle B/F}(-{\bbox{q}},-E)
  &=&
        \mp
        {
        1
        \over
         (2\pi\hbar)^3
        }
        {
        2\pi m^2
        \over
        n\beta q
        }
        {
        e^{-\beta E}
        \over
        1-
        e^{-\beta E}
        }
        \log
        \left[
        {
        1\mp z
        \exp{
        \left[
        -{
        \beta
        \over
             8m
        }
        {
        (-2mE + q^2)^2
        \over
                  q^2
        }
        \right]
        }
        \over
        1\mp z
        \exp{
        \left[
        -{
        \beta
        \over
             8m
        }
        {
        (-2mE - q^2)^2
        \over
                  q^2
        }
        \right]
        }
        }
        \right]  
        \\
  &=&
        \mp
        {
        1
        \over
         (2\pi\hbar)^3
        }
        {
        2\pi m^2
        \over
        n\beta q
        }
        {
        1
        \over
        1-
        e^{\beta E}
        }
        \log
        \left[
        {
        1\mp z
        \exp{
        \left[
        -{
        \beta
        \over
             8m
        }
        {
        (2mE + q^2)^2
        \over
                  q^2
        }
        \right]
        }
        \over
        1\mp z
        \exp{
        \left[
        -{
        \beta
        \over
             8m
        }
        {
        (2mE - q^2)^2
        \over
                  q^2
        }
        \right]
        }
        }
        \right]  
        =
        S_{\rm \scriptscriptstyle B/F}({\bbox{q}},E)
        ,
\end{eqnarray*}
which proves (\ref{dbce}).
Similarly
\begin{eqnarray*}
  e^{-\beta E}S_{\rm \scriptscriptstyle MB}(-{\bbox{q}},-E)
  &=&
        {
        1
        \over
         (2\pi\hbar)^3
        }
        {
        2\pi m^2
        \over
        n\beta q
        }
        z
         \exp\left[
        -{
        \beta E
        }
        \right]
        \exp\left[
        -{
        \beta
        \over
             8m
        }
        {
        (-2mE + q^2)^2
        \over
                  q^2
        }
        \right]
        \\
  &=&
        {
        1
        \over
         (2\pi\hbar)^3
        }
        {
        2\pi m^2
        \over
        n\beta q
        }
        z
        \exp\left[
        -{
        \beta
        \over
             8m
        }
        {
        (2mE + q^2)^2
        \over
                  q^2
        }
        \right]
        =
        S_{\rm \scriptscriptstyle MB}({\bbox{q}},E)
        .
\end{eqnarray*}
\par
The master equation (\ref{me}) for the Rayleigh gas in the case of a
free gas of Bose or Fermi particles takes therefore the form
\begin{eqnarray}
  \label{bf}
        {  
        d {\hat \varrho}  
        \over  
                      dt
        }  
        &=&
        {\cal M}_{\rm \scriptscriptstyle B/F} [\hat \varrho]
        \\
        &=&
        -
        {i \over \hbar}
        [{\hat H}_0
        ,
        {\hat \varrho}
        ]
        +
        \nonumber
        \int_{{\bf R}^3}  d\mu ({\bbox{q}}) \,  
        \Biggl[
        \hat{U} ({\bbox{q}})
        \sqrt{
        S_{\rm \scriptscriptstyle B/F}({\bbox{q}},{\hat {\bbox{p}}})
        }
        {\hat \varrho}
        \sqrt{
        S_{\rm \scriptscriptstyle B/F}({\bbox{q}},{\hat {\bbox{p}}})
        }
        \hat{U}^{\dagger} ({\bbox{q}})
        -
        \frac 12
        \left \{
        S_{\rm \scriptscriptstyle B/F}({\bbox{q}},{\hat {\bbox{p}}}),
        {\hat \varrho}
        \right \}
        \Biggr],  
\end{eqnarray}
with $S_{\rm \scriptscriptstyle B/F}({\bbox{q}},{ {\bbox{p}}})$ given
explicitly by (\ref{6}), 
and a similar expression ${\cal M}_{\rm \scriptscriptstyle MB}$ can be 
considered for a free gas of
Maxwell-Boltzmann particles.
Both ${\cal M}_{\rm \scriptscriptstyle B/F}$ and ${\cal M}_{\rm
  \scriptscriptstyle MB}$ are translation- and rotation-covariant and
admit the same stationary solution with the canonical structure
$\rho_0 (\hat{\bbox{p}})$.  
\section{BROWNIAN LIMIT}  
\label{quattro} 
In the framework of an ideal gas considered in Section~\ref{tre}, i.e.,
referring to ${\cal M}_{\rm \scriptscriptstyle B/F}$ and ${\cal M}_{\rm
  \scriptscriptstyle MB}$, we now want to consider the
physically distinguished case in which the test particle is much
heavier than the particles making up the gas, so that $\alpha=m/M$ is
much smaller than one, the so-called Brownian limit. In order to do
this we have to evaluate the dynamic structure factor for a free gas
in the case $\alpha\ll 1$. The natural starting points are expressions 
(\ref{8}) and (\ref{9}) in which the factor $\alpha$ has been put into 
evidence through the function (\ref{77}). In particular we have the
relations
\begin{eqnarray}
  \label{20}
  \frac{\beta}{2m}\sigma^2({\bbox{q}},{\bbox{p}})
  &=&
  \frac{\beta}{8m}q^2 
  + \frac{\beta}{2}\frac{1}{2M}[q^2 +2 {\bbox{p}}
  \cdot {\bbox{q}}]
  + \frac{\beta}{2}\frac{1}{q^2}\alpha\frac{1}{4M}[q^2 +2 {\bbox{p}}
  \cdot {\bbox{q}}]^2
  \nonumber \\
  &=&
  \frac{\beta}{8m}q^2 
  + \frac{\beta}{2}E ({\bbox{q}},{\bbox{p}})
  + \frac{\beta}{2}\frac{m}{q^2}E^2 ({\bbox{q}},{\bbox{p}})
  \nonumber \\
  \frac{\beta}{2m}(\sigma({\bbox{q}},{\bbox{p}})-q)^2
  &=&
  \frac{\beta}{8m}q^2 
  - \frac{\beta}{2}\frac{1}{2M}[q^2 + 2{\bbox{p}}
  \cdot {\bbox{q}}]
  + \frac{\beta}{2}\frac{1}{q^2}\alpha\frac{1}{4M}[q^2 + 2{\bbox{p}}
  \cdot {\bbox{q}}]^2
  \nonumber \\
  &=&
  \frac{\beta}{8m}q^2 
  - \frac{\beta}{2}E ({\bbox{q}},{\bbox{p}})
  + \frac{\beta}{2}\frac{m}{q^2}E^2 ({\bbox{q}},{\bbox{p}})
  \nonumber \\
  \frac{\beta}{2m}(2\sigma^2({\bbox{q}},{\bbox{p}})q-q^2 )
  &=&
  \frac{\beta}{2M}[q^2 + 2{\bbox{p}}\cdot{\bbox{q}}]
  \nonumber \\
  &=&
  \beta E ({\bbox{q}},{\bbox{p}}).
\end{eqnarray}
The Brownian limit can now be taken neglecting the terms of order
$\alpha$ in (\ref{20}) or equivalently considering small energy
transfer, corresponding to a broader time scale, and keeping in
(\ref{20}) only the terms linear in $E$, disregarding higher powers
of the energy transfer. The resulting dynamic structure factors,
denoted by an index $\infty$, are given by
        \begin{equation}
        \label{21}
        S^{\scriptscriptstyle\infty}_{\rm \scriptscriptstyle
          B/F}({\bbox{q}},{\bbox{p}}) 
        =
        \mp
        {
        1
        \over
         (2\pi\hbar)^3
        }
        {
        2\pi m^2
        \over
        n\beta q
        }
        {
        1
        \over
        1-
        e^{
        {
        \beta
        }
        \left[
        \frac{q^2}{2M}
        +\frac{{\bbox{q}}\cdot{\bbox{p}}}{M}
        \right]
        }
        }
        \log
        \left[
        {
        1\mp z
        e^{
        -{
        \beta
        \over
             8m
        }
        q^2
        }
        e^{
        -\frac{\beta}{2}
        [
        {
        q^2
        \over
             2M
        }
        +
        {{\bbox{q}}\cdot{\bbox{p}} \over M}        
        ]
        }
        \over
        1\mp z
        e^{
        -{
        \beta
        \over
             8m
        }
        q^2
        }
        e^{
        +\frac{\beta}{2}
        [
        {
        q^2
        \over
             2M
        }
        +
        {{\bbox{q}}\cdot{\bbox{p}} \over M}
        ]
        }
        }
        \right]
        \end{equation}
and
        \begin{equation}
        \label{22}
        S^{\scriptscriptstyle\infty}_{\rm \scriptscriptstyle MB}({\bbox{q}},{\bbox{p}})
        =
        {
        1
        \over
         (2\pi\hbar)^3
        }
        {
        2\pi m^2
        \over
        n\beta q
        }
        z
        e^{
        -{
        \beta
        \over
             8m
        }
        q^2
        }
        e^{
        -\frac{\beta}{2}
        [
        {
        q^2
        \over
             2M
        }
        +
        {{\bbox{q}}\cdot{\bbox{p}} \over M}
        ]
        }
        \end{equation}
respectively.
Considering the corresponding expressions
in terms of $E({\bbox{q}},{\bbox{p}})$,
        \begin{equation}
        \label{23}
        S^{\scriptscriptstyle\infty}_{\rm \scriptscriptstyle B/F}
        ({\bbox{q}},E({\bbox{q}},{\bbox{p}}))
        =
        \mp
        {
        1
        \over
         (2\pi\hbar)^3
        }
        {
        2\pi m^2
        \over
        n\beta q
        }
        {
        1
        \over
        1-
        e^{\beta E({\bbox{q}},{\bbox{p}})
        }
        }
        \log
        \left[
        {
        1\mp z
        e^{
        -{
        \beta
        \over
             8m
        }
        q^2
        }
        e^{
        -\frac{\beta}{2}
        E({\bbox{q}},{\bbox{p}})
        }
        \over
        1\mp z
        e^{
        -{
        \beta
        \over
             8m
        }
        q^2
        }
        e^{
        +\frac{\beta}{2}
        E({\bbox{q}},{\bbox{p}})
        }
        }
        \right]
\end{equation}
and
\begin{equation}
  \label{24}
        S^{\scriptscriptstyle\infty}_{\rm \scriptscriptstyle MB}
        ({\bbox{q}},E({\bbox{q}},{\bbox{p}}))
        =
        {
        1
        \over
         (2\pi\hbar)^3
        }
        {
        2\pi m^2
        \over
        n\beta q
        }
        z
        e^{
        -{
        \beta
        \over
             8m
        }
        q^2
        }
        e^{
        -\frac{\beta}{2}
        E({\bbox{q}},{\bbox{p}})
        }
      ,
        \end{equation}
one immediately sees that rotational invariance is preserved in this
approximation. One can check that also the detailed balance condition
is not spoiled, in fact from (\ref{23}) one has
\begin{eqnarray*}
        e^{-\beta E}S^{\scriptscriptstyle\infty}_{\rm \scriptscriptstyle B/F}
        (-{\bbox{q}},-E)
        &=&
        \mp
        {
        1
        \over
         (2\pi\hbar)^3
        }
        {
        2\pi m^2
        \over
        n\beta q
        }
        {
        e^{-\beta E}
        \over
        1-
        e^{-\beta E
        }
        }
        \log
        \left[
        {
        1\mp z
        e^{
        -{
        \beta
        \over
             8m
        }
        q^2
        }
        e^{
        +\frac{\beta}{2}
        E
        }
        \over
        1\mp z
        e^{
        -{
        \beta
        \over
             8m
        }
        q^2
        }
        e^{
        -\frac{\beta}{2}
        E
        }
        }
        \right]
        \\
        &=&
        \mp
        {
        1
        \over
         (2\pi\hbar)^3
        }
        {
        2\pi m^2
        \over
        n\beta q
        }
        {
        1
        \over
        1-
        e^{\beta E
        }
        }
        \log
        \left[
        {
        1\mp z
        e^{
        -{
        \beta
        \over
             8m
        }
        q^2
        }
        e^{
        -\frac{\beta}{2}
        E
        }
        \over
        1\mp z
        e^{
        -{
        \beta
        \over
             8m
        }
        q^2
        }
        e^{
        +\frac{\beta}{2}
        E
        }
        }
        \right]
        =
        S^{\scriptscriptstyle\infty}_{\rm \scriptscriptstyle B/F}
        ({\bbox{q}},E)
\end{eqnarray*}
and from (\ref{24})
\begin{displaymath}
        e^{-\beta E}  S^{\scriptscriptstyle\infty}_{\rm \scriptscriptstyle MB}
        (-{\bbox{q}},-E)
        =
        {
        1
        \over
         (2\pi\hbar)^3
        }
        {
        2\pi m^2
        \over
        n\beta q
        }
        z
        e^{
        -{
        \beta
        \over
             8m
        }
        q^2
        }
        e^{
        -{\beta}
        E({\bbox{q}},{\bbox{p}})
        }
        e^{
        \frac{\beta}{2}
        E({\bbox{q}},{\bbox{p}})
        }
      =
      S^{\scriptscriptstyle\infty}_{\rm \scriptscriptstyle MB}
        ({\bbox{q}},E)
        .
\end{displaymath}
As a result in place of (\ref{bf}) we now consider the mapping ${\cal
  M}^{\scriptscriptstyle\infty}_{\rm \scriptscriptstyle B/F}$ 
\begin{eqnarray}
  \label{bfinfty}
        {  
        d {\hat \varrho}  
        \over  
                      dt
        }  
        &=&
        {\cal M}^{\scriptscriptstyle\infty}_{\rm \scriptscriptstyle B/F} 
        [\hat \varrho]
        \\
        &=&
        -
        {i \over \hbar}
        [{\hat H}_0
        ,
        {\hat \varrho}
        ]
        +
        \nonumber
        \int_{{\bf R}^3}  d\mu ({\bbox{q}}) \,  
        \Biggl[
        \hat{U} ({\bbox{q}})
        \sqrt{
        S^{\scriptscriptstyle\infty}_{\rm \scriptscriptstyle B/F}
        ({\bbox{q}},{\hat {\bbox{p}}})
        }
        {\hat \varrho}
        \sqrt{
        S^{\scriptscriptstyle\infty}_{\rm \scriptscriptstyle B/F}
        ({\bbox{q}},{\hat {\bbox{p}}})
        }
        \hat{U}^{\dagger} ({\bbox{q}})
        -
        \frac 12
        \left \{
        S^{\scriptscriptstyle\infty}_{\rm \scriptscriptstyle B/F}
        ({\bbox{q}},{\hat {\bbox{p}}}),
        {\hat \varrho}
        \right \}
        \Biggr]  
        ,
\end{eqnarray}
ans similarly ${\cal M}^{\scriptscriptstyle\infty}_{\rm
  \scriptscriptstyle MB}$ for Maxwell-Boltzmann statistics.
${\cal M}^{\scriptscriptstyle\infty}_{\rm \scriptscriptstyle B/F}$ and 
${\cal M}^{\scriptscriptstyle\infty}_{\rm \scriptscriptstyle MB}$  
are still translation- and rotation-covariant and
admit the same stationary solution with the canonical structure
$\rho_0 (\hat{\bbox{p}})$.  
\par
In the master equation (\ref{me}), or according to the physical system 
under consideration (\ref{bf}) or (\ref{bfinfty}), the quantum
scattering rate or transition probability  
appears through the dynamic structure factor and the square modulus of 
the Fourier transform of the T matrix determining the integration
measure (\ref{misura}), these quantities being connected to the
differential scattering cross-section by (\ref{diff}). In order to
pass from the master equation to the related Fokker-Planck equation
through a Kramers-Moyal expansion, as stressed by van
Kampen~\cite{vanKampen} we need to put into evidence a small parameter 
governing the size of the fluctuations in the macroscopic system. In
our case this parameter is naturally given by the momentum transfer
${\bbox{q}}$, which through the dynamic structure factor is directly
linked to the equilibrium fluctuations of the macroscopic
system. Small ${\bbox{q}}$ means long-wavelength fluctuations,
corresponding to the macroscopic, long range properties of the
environment.
It is physically meaningful to consider
both approximations $|{\bbox{q}}|\ll 1$ and $\alpha\ll 1$, or
equivalently small energy transfer, together, so that starting from
the Maxwell-Boltzmann case
\begin{eqnarray*}
        {\cal M}^{\scriptscriptstyle\infty}_{\rm \scriptscriptstyle MB} [\cdot]
        =
        &-&
        {i \over \hbar}
        [{\hat H}_0
        ,
        \cdot
        ]
        +
        z
        {4\pi^2 m^2 \over\beta\hbar}
        \int_{{\bf R}^3}  d^3\!
        {\bbox{q}}
        \,  
        {
        | \tilde{t} (q) |^2
        \over
        q
        }
        e^{-
        {
        \beta
        \over
             8m
        } (1+2\alpha)
        {{{q}}^2}
        }
      \\
      &&
      \hphantom{
        z
        {4\pi^2 m^2 \over\beta\hbar}
        \int_{{\bf R}^3}  d^3\!
        {\bbox{q}}
        \,  
        {
        | \tilde{t} (q) |^2
        \over
        q
        }
        }
      \times
        \Biggl[
        e^{{i\over\hbar}{\bbox{q}}\cdot{\hat {\bbox{x}}}}
        e^{-{\beta\over 4M}{\bbox{q}}\cdot{\hat {\bbox{p}}}}
        \cdot
        e^{-{\beta\over 4M}{\bbox{q}}\cdot{\hat {\bbox{p}}}}
        e^{-{i\over\hbar}{\bbox{q}}\cdot{\hat {\bbox{x}}}}
        - {1\over 2}
        \left \{
        e^{-{\beta\over 2M}{\bbox{q}}\cdot{\hat {\bbox{p}}}}
        ,
        \cdot
        \right \}
        \Biggr]
\end{eqnarray*}
we expand the dissipative part of the mapping in the small parameter
${\bbox{q}}$. We will expand the operators depending on ${\bbox{q}}$
up to second order, so as to have contributions at most bilinear in
the operators ${\hat {\bbox{x}}}$ and ${\hat {\bbox{p}}}$, position
and momentum of the Brownian particle. We thus obtain a structure
analogous to the classical Fokker-Planck equation, with a friction
term linearly proportional to velocity: this class of models is known
as quantum Brownian
motion~\cite{AlbertoQBM,LindbladQBM,IsarJMP-LindbladJMP}. Recalling that
$\alpha\ll 1$ and exploiting the symmetry properties of the
integration measure the result for the dissipative part is~\cite{art3}
        \begin{displaymath}
        -
        z
        {2\pi^2 m^2 \over\beta\hbar}
        \int_{{\bf R}^3} d^3\!
        {\bbox{q}}
        \,  
        {
        | \tilde{t} (q) |^2
        \over
        q
        }
        e^{-
        {
        \beta
        \over
             8m
        }
        {{{q}}^2}
        }
        \sum_{i=1}^3
        {\bbox{q}}^2_i
        \biggl\{
        {
        1
        \over
        \hbar^2
        }
        \left[
        {\hat{\bbox{x}}}_i ,
        \left[
        {\hat{\bbox{x}}}_i , \cdot
        \right]
        \right]
        +
        {
        \beta^2
        \over
         16 M^2
        }
        \left[
         {\hat{\bbox{p}}}_i  ,
        \left[
         {\hat{\bbox{p}}}_i  , \cdot
        \right]
        \right]
        +
        {i\over\hbar}
        {
        \beta
        \over
        2M
        }
        \left[
        {\hat{\bbox{x}}}_i ,
        \left \{
         {\hat{\bbox{p}}}_i  , \cdot
        \right \}
        \right]
        \biggr\}
        \end{displaymath}
where $i=1,2,3$ denotes Cartesian coordinates. Due to isotropy of the
environment we have $
{\bbox{q}}^2_i =
\frac 13 q^2
$, so that we can define the
coefficients
        \begin{eqnarray}
        \label{coeff}  
        D_{pp}
        &=&
        \frac 23
        {\pi^2 m^2 \over\beta\hbar}
        \int_{{\bf R}^3} d^3\!
        {\bbox{q}}
        \,  
        {
        | \tilde{t} (q) |^2
        }
        q
        e^{-
        {
        \beta
        \over
             8m
        }
        {{{q}}^2}
        }
        \nonumber
        \\
        D_{xx} 
        &=& 
        \left(
        {
        \beta\hbar
        \over
            4M
        }\right)^2 D_{pp}
        \\
        \gamma 
        &=&
        \left({
        \beta
        \over
             2M
        }\right)
        D_{pp}
        \nonumber
        \end{eqnarray}
and introduce the following mapping describing quantum dissipation
\begin{equation}
  \label{qd}
  {\cal L}_{\rm \scriptscriptstyle QD}[\cdot]=
        -
        {
        D_{pp}  
        \over
         \hbar^2
        }
        \sum_{i=1}^3
        \left[  
        {\hat{\bbox{x}}}_i ,
        \left[  
        {\hat{\bbox{x}}}_i ,\cdot
        \right]  
        \right]  
        -
        {
        D_{xx}
        \over
         \hbar^2
        }
        \sum_{i=1}^3
        \left[  
         {\hat{\bbox{p}}}_i ,
        \left[  
         {\hat{\bbox{p}}}_i ,\cdot
        \right]  
        \right]  
        -
        {i\over\hbar}
        \gamma
        \sum_{i=1}^3
        \left[  
        {\hat{\bbox{x}}}_i ,
        \left \{  
         {\hat{\bbox{p}}}_i ,\cdot
        \right \}  
        \right]      
        ,
\end{equation}
thus coming to  the Fokker-Planck equation
\begin{equation}
  \label{lmb}
        {  
        d {\hat \varrho}  
        \over  
                      dt
        }  
      =
        -
        {i \over \hbar}
        [{\hat H}_0
        ,
        {\hat \varrho}
        ]
        +
        {\cal L}_{\rm \scriptscriptstyle MB} [\hat \varrho]
      =
        -
        {i \over \hbar}
        [{\hat H}_0
        ,
        {\hat \varrho}
        ]
        +
        z{\cal L}_{\rm \scriptscriptstyle QD} [\hat \varrho]
        .
\end{equation}
\par
In view of the result (\ref{lmb}) for the Fokker-Planck equation
describing the motion of the Brownian particle in a gas obeying
Maxwell-Boltzmann statistics, we now look for the corrections to
(\ref{lmb}) brought about by quantum statistics at finite
temperature. As usual we will deal with both Bose and Fermi
statistics, exploiting expression (\ref{b8}) obtained in
Appendix~\ref{bb} by deriving an exact expansion for 
$S^{\scriptscriptstyle\infty}_{\rm \scriptscriptstyle B/F}
        ({\bbox{q}},E)$: 
\begin{eqnarray}
  \label{40}
        S^{\scriptscriptstyle\infty}_{\rm \scriptscriptstyle B/F}
        ({\bbox{q}},E)
        &=&
        S^{\scriptscriptstyle\infty}_{\rm \scriptscriptstyle MB}
        ({\bbox{q}},E)
        \left[
          \sum_{k=0}^{\infty} (\pm z)^k -
          \frac{\beta}{8m} q^2\sum_{k=1}^{\infty}(\pm)^k k z^k
          \right.
          \\
          &&
          \hphantom{{S^{\scriptscriptstyle\infty}_{\rm \scriptscriptstyle MB}
        ({\bbox{q}},E)
          \sum_{k=0}^{\infty} 
          }}
        \left.
          + \frac{1}{12} (\beta E)^2\sum_{k=1}^{\infty}(\pm)^k k z^k
          + \frac{1}{24} (\beta E)^2\sum_{k=1}^{\infty}(\pm)^k k^2 z^k
          + {\mathord{\mathrm{O}}} (q^4)
        \right]
        ,
        \nonumber
\end{eqnarray}
where a suitable expansion in the small parameter ${\bbox{q}}$ has
already been performed. 
Let us first introduce the Bose-Einstein and the Fermi-Dirac
functions~\cite{Pathria,Ryzhik}, given by
\begin{equation}
  \label{gg} 
  g_\nu (z) = 
  \frac{1}{\Gamma (\nu)}
  \int_0^{+\infty} dx \, \frac{x^{\nu -1}}{z^{-1}e^x -1}
  \qquad 0\leq z <1
  , 
  \>
  \nu> 0
\end{equation}
and 
\begin{equation}
  \label{ff} 
  f_\nu (z) = 
  \frac{1}{\Gamma (\nu)}
  \int_0^{+\infty} dx \, \frac{x^{\nu -1}}{z^{-1}e^x +1}
  \qquad   0\leq z<\infty
  , 
  \> \nu> 0
\end{equation} 
respectively, related for integer $\nu$ by $f_n (z)=-g_n (-z)$.
These functions, typically appearing in the quantum statistical
mechanics of Bose and Fermi systems, satisfy the recurrence relations
\begin{eqnarray}
\label{ricorre}
  g_{\nu -1} (z)=z \frac{\partial}{\partial z}[g_{\nu} (z)]
  \qquad
  f_{\nu -1} (z)=z \frac{\partial}{\partial z}[f_{\nu} (z)]
  ,
\end{eqnarray}
so that they can be defined also for $\nu\leq 0$. Starting from
(\ref{ricorre}) and exploiting the following representations for $|z|<1$
\begin{eqnarray*}
  g_\nu (z) =\sum_{k=1}^{\infty} \frac{z^k}{k^\nu} 
  \qquad
  f_\nu (z) =\sum_{k=1}^{\infty} (-)^{k-1}  \frac{z^k}{k^\nu} 
  ,
\end{eqnarray*}
one can write (\ref{40}) in the Bose case as
\begin{equation}
  \label{41}
        S^{\scriptscriptstyle\infty}_{\rm \scriptscriptstyle B}
        ({\bbox{q}},E)
        =
        S^{\scriptscriptstyle\infty}_{\rm \scriptscriptstyle MB}
        ({\bbox{q}},E)
        \left[
          {g_0 (z) \over z}
          - \frac{\beta}{8m} q^2 g_{-1} (z)
          + \frac{1}{24} (\beta E)^2 (g_{-2} (z)+2g_{-1} (z))
          + {\mathord{\mathrm{O}}} (q^4)
        \right]
\end{equation}
and in the Fermi case as
\begin{equation}
  \label{43}
        S^{\scriptscriptstyle\infty}_{\rm \scriptscriptstyle F}
        ({\bbox{q}},E)
        =
        S^{\scriptscriptstyle\infty}_{\rm \scriptscriptstyle MB}
        ({\bbox{q}},E)
        \left[
          {f_0 (z) \over z}
          +\frac{\beta}{8m} q^2 f_{-1} (z)
          -\frac{1}{24} (\beta E)^2 (f_{-2} (z)+2f_{-1} (z))
          + {\mathord{\mathrm{O}}} (q^4)
        \right],
\end{equation}
where the functions appearing in (\ref{41}) can be written for $0\leq
z<1$ in closed form as:
\begin{eqnarray}
  \label{42}
  g_0 (z)&=&{z\over 1-z}
  \\
  g_{-1} (z)&=&{z\over (1-z)^2}
  \qquad
  g_{-2} (z)={z+z^2\over (1-z)^3}
  \nonumber
  ,
\end{eqnarray}
while the functions appearing in (\ref{43}) can be written for $0\leq
z<\infty$ in closed form as:
\begin{eqnarray}
  \label{44}
  f_0 (z)&=&{z\over 1+z}
  \\
  f_{-1} (z)&=&{z\over (1+z)^2}
  \qquad
  f_{-2} (z)={z-z^2\over (1+z)^3}
  \nonumber
  .
\end{eqnarray}
To evaluate the corrections due to quantum statistics we note that
when 
$
        S^{\scriptscriptstyle\infty}_{\rm \scriptscriptstyle MB}
        ({\bbox{q}},E({\bbox{q}},{\bbox{p}}))
$
is substituted by an expression of the form
\begin{displaymath}
          S^{\scriptscriptstyle\infty}_{\rm \scriptscriptstyle MB}
        ({\bbox{q}},E({\bbox{q}},{\bbox{p}}))
        A[1+2B q^2+2C ({\bbox{p}}\cdot{\bbox{q}})^2 ],
\end{displaymath}
so that keeping terms at most quadratic in ${\bbox{q}}$ in the correction
\begin{displaymath}
  \sqrt{
          S^{\scriptscriptstyle\infty}_{\rm \scriptscriptstyle MB}
        ({\bbox{q}},E({\bbox{q}},{\bbox{p}}))
       }
     \rightarrow
  \sqrt{
          S^{\scriptscriptstyle\infty}_{\rm \scriptscriptstyle MB}
        ({\bbox{q}},E({\bbox{q}},{\bbox{p}}))
       }
     \sqrt{A}
       [1+B q^2+C ({\bbox{p}}\cdot{\bbox{q}})^2 ],
\end{displaymath}
the mapping ${\cal M}^{\scriptscriptstyle\infty}_{\rm
  \scriptscriptstyle MB}$ always in the same approximation becomes
\begin{eqnarray}
  \label{45}
        &-&
        {i \over \hbar}
        [{\hat H}_0
        ,
        \cdot
        ]
        \\
        &+&
        A
        \int_{{\bf R}^3}  d\mu ({\bbox{q}}) \,  
        \Biggl[
        \hat{U} ({\bbox{q}})
        \sqrt{
        S^{\scriptscriptstyle\infty}_{\rm \scriptscriptstyle MB}
        ({\bbox{q}},{\hat {\bbox{p}}})
        }
        \cdot
        \sqrt{
        S^{\scriptscriptstyle\infty}_{\rm \scriptscriptstyle MB}
        ({\bbox{q}},{\hat {\bbox{p}}})
        }
        \hat{U}^{\dagger} ({\bbox{q}})
        -
        \frac 12
        \left \{
        S^{\scriptscriptstyle\infty}_{\rm \scriptscriptstyle MB}
        ({\bbox{q}},{\hat {\bbox{p}}}),
        \cdot
        \right \}
        \Biggr]
        \nonumber
        \\
        &+&
        2AB
        \int_{{\bf R}^3}  d\mu ({\bbox{q}}) \,  q^2
        \Biggl[
        \hat{U} ({\bbox{q}})
        \sqrt{
        S^{\scriptscriptstyle\infty}_{\rm \scriptscriptstyle MB}
        ({\bbox{q}},{\hat {\bbox{p}}})
        }
        \cdot
        \sqrt{
        S^{\scriptscriptstyle\infty}_{\rm \scriptscriptstyle MB}
        ({\bbox{q}},{\hat {\bbox{p}}})
        }
        \hat{U}^{\dagger} ({\bbox{q}})
        -
        \frac 12
        \left \{
        S^{\scriptscriptstyle\infty}_{\rm \scriptscriptstyle MB}
        ({\bbox{q}},{\hat {\bbox{p}}}),
        \cdot
        \right \}
        \Biggr]
        \nonumber
        \\
        &+&
        AC
        \int_{{\bf R}^3}  d\mu ({\bbox{q}}) \,  
        \Biggl[
        \hat{U} ({\bbox{q}})
        \sqrt{
        S^{\scriptscriptstyle\infty}_{\rm \scriptscriptstyle MB}
        ({\bbox{q}},{\hat {\bbox{p}}})
        }
        \{({\hat {\bbox{p}}}\cdot{\bbox{q}})^2 ,\cdot\}
        \sqrt{
        S^{\scriptscriptstyle\infty}_{\rm \scriptscriptstyle MB}
        ({\bbox{q}},{\hat {\bbox{p}}})
        }
        \hat{U}^{\dagger} ({\bbox{q}})
        -
        \left \{
        S^{\scriptscriptstyle\infty}_{\rm \scriptscriptstyle MB}
        ({\bbox{q}},{\hat {\bbox{p}}})({\hat {\bbox{p}}}\cdot{\bbox{q}})^2,
        \cdot
        \right \}
        \Biggr]
        \nonumber
        .
\end{eqnarray}
Looking at (\ref{45}) one immediately sees that, keeping terms at most
quadratic in ${\bbox{q}}$, the last two terms are to be neglected,
since the dynamic structure factor 
$S^{\scriptscriptstyle\infty}_{\rm \scriptscriptstyle MB}
        ({\bbox{q}},{\hat {\bbox{p}}})$
and the unitary operators $\hat{U} ({\bbox{q}})$ can now only bring in 
a constant factor. The only change in the structure of the mapping is
therefore given by the numerical factor $A$ multiplying the
dissipative part. This factor is actually given by
${g_0 (z) / z}$ in the Bose case and by ${f_0 (z) / z}$ in the 
Fermi case. The Fokker-Planck equation (\ref{lmb}) in the case of Bose 
statistics of the gas therefore becomes
\begin{equation}
  \label{lbe}
        {  
        d {\hat \varrho}  
        \over  
                      dt
        }  
      =
        -
        {i \over \hbar}
        [{\hat H}_0
        ,
        {\hat \varrho}
        ]
        +
        {\cal L}_{\rm \scriptscriptstyle B} [\hat \varrho]
      =
        -
        {i \over \hbar}
        [{\hat H}_0
        ,
        {\hat \varrho}
        ]
        +
        g_0 (z) {\cal L}_{\rm \scriptscriptstyle QD} [\hat \varrho]
        ,
\end{equation}
while for Fermi particles one has
\begin{equation}
  \label{lfd}
        {  
        d {\hat \varrho}  
        \over  
                      dt
        }  
      =
        -
        {i \over \hbar}
        [{\hat H}_0
        ,
        {\hat \varrho}
        ]
        +
        {\cal L}_{\rm \scriptscriptstyle F} [\hat \varrho]
      =
        -
        {i \over \hbar}
        [{\hat H}_0
        ,
        {\hat \varrho}
        ]
        +
        f_0 (z) {\cal L}_{\rm \scriptscriptstyle QD} [\hat \varrho]
        ,
\end{equation}
and the following simple relations hold
\begin{eqnarray}
  \label{50}
  {\cal L}_{\rm \scriptscriptstyle MB}
  &=&z{\cal L}_{\rm \scriptscriptstyle QD}
  \nonumber
  \\
  {\cal L}_{\rm \scriptscriptstyle B}
  &=& g_0 (z){\cal L}_{\rm \scriptscriptstyle QD}
  = {z\over 1-z}{\cal L}_{\rm \scriptscriptstyle QD}
  = {1\over 1-z}{\cal L}_{\rm \scriptscriptstyle MB}
  \\
  {\cal L}_{\rm \scriptscriptstyle F}
  &=& f_0 (z){\cal L}_{\rm \scriptscriptstyle QD}
  = {z\over 1+z}{\cal L}_{\rm \scriptscriptstyle QD}
  = {1\over 1+z}{\cal L}_{\rm \scriptscriptstyle MB}
  .
  \nonumber
\end{eqnarray}
According to (\ref{50}) and setting after (\ref{coeff})
\begin{equation}
  \label{51}
        \gamma_{\rm \scriptscriptstyle MB}=z\gamma=
        z
        {
        \beta
        \over
             2M
        }
        D_{pp}
        =
        z
        \frac 13
        {\pi^2 m^2 \over M\hbar}
        \int d^3\!
        {\bbox{q}}
        \,  
        {
        | \tilde{t} (q) |^2
        }
        q
        e^{-
        {
        \beta
        \over
             8m
        }
        {{{q}}^2}
        }
\end{equation}
one has the following very simple relation between the friction
coefficients in (\ref{lmb}) and  (\ref{lbe}) or  (\ref{lfd}):
\begin{equation}
        \label{rapporto}
        \gamma_{\rm \scriptscriptstyle B/F}=
        {
        \gamma_{\rm \scriptscriptstyle MB}
        \over
         1\mp z
        }
      .
\end{equation}
The relationship between the Fokker-Planck equations for
Maxwell-Boltzmann or Bose and Fermi statistics, as given respectively
by (\ref{lmb}),  (\ref{lbe}) and  (\ref{lfd}), is actually remarkably
simple: they have the very same operator structure, apart from an
overall coefficient depending on the fugacity of the gas, which
determines the relative weight of the dissipative contribution to the
dynamics. As it is to be expected only the statistics of the reservoir 
is of relevance, since the test particle is a single particle. The
Fokker-Planck equations obtained for the description of quantum
dissipation may be compactly written:
\begin{equation}
  \label{compatto}
        {  
        d {\hat \varrho}  
        \over  
                      dt
        }  
      =
        -
        {i \over \hbar}
        [{\hat H}_0
        ,
        {\hat \varrho}
        ]
        +
        \zeta (z) {\cal L}_{\rm \scriptscriptstyle QD} [\hat \varrho]
\end{equation}
with $\zeta (z)$ defined as follows
\begin{equation}
  \label{zeta}
  \zeta (z)=
   \left\{ \matrix{ z & \qquad {\rm Maxwell-Boltzmann}  \cr 
       z/ (1-z) & {\rm Bose} \cr z/ (1+z)  & {\rm Fermi} }\right.
   .
\end{equation}
\par
We now briefly come back to the question dealt with at the end of
Section~\ref{due} about the physical relevance of the structure of
translation-covariant master equation obtained in~\cite{Manita}. As
already stressed the master equation (\ref{me}), while having the
general translation-covariant structure considered
in~\cite{HolevoTI}, does not comply with the further restrictions
given in~\cite{Manita}, while this would be the case if instead of the 
dynamic structure factor $S ({\bbox{q}},{\bbox{p}})$ one would
consider the symmetrized correlation function
$\tilde{S}({\bbox{q}},{\bbox{p}})$, which is an even function of
${\bbox{q}}$ and $E({\bbox{q}},{\bbox{p}})$. This could be considered
a natural phenomenological Ansatz in view of the
result~\cite{Manita}. In Section~\ref{due} we showed however that this
substitution would spoil the existence of the expected stationary
solution. More than this, if we now consider the Brownian limit, the
symmetrized version of 
$S^{\scriptscriptstyle\infty}_{\rm \scriptscriptstyle MB}
        ({\bbox{q}},{\hat {\bbox{p}}})$, which can be immediately
obtained from (\ref{24}), reads
\begin{displaymath}
        \tilde{S}^{\scriptscriptstyle\infty}_{\rm \scriptscriptstyle MB}
        ({\bbox{q}},E({\bbox{q}},{\bbox{p}}))
        =
        {
        1
        \over
         (2\pi\hbar)^3
        }
        {
        2\pi m^2
        \over
        n\beta q
        }
        z
        e^{
        -{
        \beta
        \over
             8m
        }
        q^2
        }
      ,
\end{displaymath}
so that the dependence on ${\bbox{p}}$ is completely lost and the
whole operator structure in (\ref{qd}) and (\ref{lmb}) is washed out,
apart from the double commutator with the position operators
${\hat{\bbox{x}}_i}$. In particular the friction term is missing, so that,
even though a Lindblad structure is retained, only a completely
different physics can be described. In the same way, for the Bose or
Fermi dynamic structure factor in the Brownian limit one has from
(\ref{b8})
\begin{eqnarray*}
        \tilde{S}^{\scriptscriptstyle\infty}_{\rm \scriptscriptstyle B/F}
        ({\bbox{q}},E)
        &=&
        \tilde{S}^{\scriptscriptstyle\infty}_{\rm \scriptscriptstyle MB}
        ({\bbox{q}},E)
        \left[
          \sum_{k=0}^{\infty} (\pm z)^k -
          \frac{\beta}{8m} q^2\sum_{k=1}^{\infty}(\pm)^k k z^k
          \right.
          \\
          &&
          \hphantom{{S^{\scriptscriptstyle\infty}_{\rm \scriptscriptstyle MB}
        ({\bbox{q}},E)
          \sum_{k=0}^{\infty} 
          }}
        \left.
          + \frac{1}{12} (\beta E)^2\sum_{k=1}^{\infty}(\pm)^k k z^k
          + \frac{1}{24} (\beta E)^2\sum_{k=1}^{\infty}(\pm)^k k^2 z^k
          + {\mathord{\mathrm{O}}} (q^4)
        \right]
        \nonumber
\end{eqnarray*}
and once again, recalling (\ref{45}) written in terms of
$\tilde{S}^{\scriptscriptstyle\infty}_{\rm \scriptscriptstyle MB}$
rather than ${S}^{\scriptscriptstyle\infty}_{\rm \scriptscriptstyle
  MB}$, one sees that under the same approximations as before the
operator structure in the dissipative part of both (\ref{lbe}) and
(\ref{lfd}) is washed out apart from the contribution due
to the double commutator in the position operators of the particle
${\hat{\bbox{x}}_i}$. 
\par
We now consider some structural features of the 
mapping ${\cal L}_{\rm \scriptscriptstyle QD}$ given by
(\ref{qd}) in terms of which the Fokker-Planck equation
(\ref{compatto}) encompassing all three statistics is given. 
$G$-covariance of ${\cal L}_{\rm \scriptscriptstyle QD}$ under
translations and rotations immediately follows from its very structure 
and the transformation laws for the operators ${\hat {\bbox{x}}}$ 
and ${\hat {\bbox{p}}}$:
\begin{eqnarray*}
  \hat{U}^{\dagger} ({\bbox{a}}){\hat {\bbox{x}}}\hat{U}
  ({\bbox{a}})={\hat {\bbox{x}}}+{\bbox{a}},
  \quad
  \hat{U}^{\dagger} ({\bbox{a}}){\hat {\bbox{p}}}\hat{U}
  ({\bbox{a}})={\hat {\bbox{p}}},
  \quad
  \hat{U}^{\dagger} (R){\hat {\bbox{x}}}\hat{U}
  (R)=R{\hat {\bbox{x}}},
  \quad
  \hat{U}^{\dagger} (R){\hat {\bbox{p}}}\hat{U}
  (R)=R{\hat {\bbox{p}}}
  .
\end{eqnarray*}
One can also see that an operator with the expected
canonical structure is a stationary solution of (\ref{compatto}) in
that
\begin{displaymath}
{\cal L}_{\rm
  \scriptscriptstyle QD}[\rho_0 ({\hat {\bbox{p}}})]=0
  ,
\end{displaymath}
due to the relationship
\begin{equation}
  \label{equilibrio}
  \frac{\gamma}{D_{pp}}=\frac{\beta}{2M}
\end{equation}
obeyed by the coefficients defined in (\ref{coeff}) and entering in
(\ref{qd}). 
A few more remarks are in order. The typical
structure of translation-covariant mappings describing quantum
dissipation in analogy with the classical Fokker-Planck equation that
one finds in the physical literature is given by~\cite{Sandulescu}
\begin{equation}
  \label{fp}
 {\cal L}^{\chi}_{\rm
  \scriptscriptstyle FP}[\cdot]
=
       -
        {i\over\hbar}
        \gamma
        \sum_{i=1}^3
        \left[  
        {\hat{\bbox{x}}}_i ,
        \left \{  
         {\hat{\bbox{p}}}_i ,\cdot
        \right \}  
        \right]      
        -
        {
        1
        \over
         \hbar^2
        }
      \frac{2M\gamma}{\beta}
        \sum_{i=1}^3
        \left[  
        {\hat{\bbox{x}}}_i ,
        \left[  
        {\hat{\bbox{x}}}_i ,\cdot
        \right]  
        \right]  
        -
        \chi
        \frac{\beta\gamma}{M}
        \sum_{i=1}^3
        \left[  
         {\hat{\bbox{p}}}_i ,
        \left[  
         {\hat{\bbox{p}}}_i ,\cdot
        \right]  
        \right]  
        ,
\end{equation}
where the ratio between the first two coefficients, given by
$\beta/2M$ as in (\ref{equilibrio}) is fixed by the requirement that
$\rho_0 ({\hat {\bbox{p}}})$ be a stationary solution, i.e., 
$
 {\cal L}^{\chi}_{\rm
  \scriptscriptstyle FP}[\rho_0 ({\hat {\bbox{p}}})]=0
$, and the only freedom left, apart from the overall multiplying
coefficient $\gamma$ is given by the adimensional factor $\chi$. If
one further asks that (\ref{fp}) can be cast in Lindblad form, so that 
$
 {\cal L}^{\chi}_{\rm
  \scriptscriptstyle FP}
$ is the generator of a completely positive dynamical 
semigroup~\cite{LindbladQBM}, one has 
the further simple requirement~\cite{znf,AlbertoQBM}
\begin{equation}
  \label{vincolo}
  \chi\geq \frac{1}{8}
.
\end{equation}
In fact under this condition, observing that for the operators
\begin{displaymath}
  {\hat{\bbox{B}}}_i{}_\pm = {\hat{\bbox{x}}}_i \pm i\kappa {\hat{\bbox{p}}}_i 
\end{displaymath}
we have the identity
\begin{displaymath}
 {\hat{\bbox{B}}}_i{}_\pm \cdot {\hat{\bbox{B}}}^{\dagger}_i{}_\pm  - \frac{1}{2} 
 \{ {\hat{\bbox{B}}}^{\dagger}_i{}_\pm {\hat{\bbox{B}}}_i{}_\pm , \cdot \}
=
-
\frac{1}{2}
\{
        \left[  
        {\hat{\bbox{x}}}_i ,
        \left[  
        {\hat{\bbox{x}}}_i ,\cdot
        \right]  
        \right]  
+
\kappa^2
        \left[  
         {\hat{\bbox{p}}}_i ,
        \left[  
         {\hat{\bbox{p}}}_i ,\cdot
        \right]  
        \right]  
\pm
2i
\kappa
        \left[  
        {\hat{\bbox{x}}}_i ,
        \left \{  
         {\hat{\bbox{p}}}_i ,\cdot
        \right \}  
        \right]      
\mp
i
\kappa
        \left[  
        \left \{  
        {\hat{\bbox{x}}}_i , {\hat{\bbox{p}}}_i 
        \right \}  
         ,
         \cdot
        \right]      
\}
,
\end{displaymath}
we may write ${\cal L}^{\chi}_{\rm
  \scriptscriptstyle FP}$ in an explicit Lindblad form in terms of the two 
generators
\begin{displaymath}
  {\hat {\bbox{L}}}_i{}_{+}=  {\hat{\bbox{x}}}_i + i\frac{\hbar
    \beta}{M}\sqrt{\frac{\chi}{2}} {\hat{\bbox{p}}}_i , 
  \qquad
  {\hat {\bbox{L}}}_i{}_{-}=  {\hat{\bbox{x}}}_i - i\frac{\hbar
    \beta}{M}\sqrt{\frac{\chi}{2}} {\hat{\bbox{p}}}_i 
\end{displaymath}
according to:
\begin{eqnarray}
  \label{fpgenerica}
  {\cal L}^{\chi}_{\rm
  \scriptscriptstyle FP}[\cdot]
  =
  &+& {2\gamma M\over \hbar^2 \beta}
  \left(
  1+\sqrt{\frac{1}{8\chi}}
  \right)
        \sum_{i=1}^3
  \left[
  {\hat {\bbox{L}}}_i{}_{+}\cdot{\hat {\bbox{L}}}_i^{\dagger}{}_{+} - 
  \frac{1}{2}\{{\hat {\bbox{L}}}^{\dagger}_i{}_{+}{\hat {\bbox{L}}}_i{}_{+} ,\cdot\} 
  \right]
  \nonumber
  \\
  &+&
  {2\gamma M\over \hbar^2 \beta}
  \left(
  1-\sqrt{\frac{1}{8\chi}}
  \right)
        \sum_{i=1}^3
  \left[
  {\hat {\bbox{L}}}_i{}_{-}\cdot{\hat {\bbox{L}}}_i^{\dagger}{}_{-} -
  \frac{1}{2}\{{\hat {\bbox{L}}}^{\dagger}_i{}_{-}{\hat {\bbox{L}}}_i{}_{-} ,\cdot\}   
  \right]
  \nonumber
  \\
  &-&
  \frac{i}{\hbar}\frac{\gamma}{2}
        \sum_{i=1}^3
        \left[  
        \left \{  
        {\hat{\bbox{x}}}_i , {\hat{\bbox{p}}}_i 
        \right \}  
         ,
         \cdot
        \right]      
.
\end{eqnarray}
The Fokker-Planck structure ${\cal L}_{\rm \scriptscriptstyle QD}$
falls within this class, with the coefficient $\gamma$ given by
(\ref{coeff}) in terms of a
suitable integral of the Fourier transform of the T matrix describing
collisions at microphysical level. Moreover it corresponds to the
value $\chi =1/8$ in (\ref{vincolo}), so that
\begin{equation}
  \label{otto}
  {\cal L}_{\rm \scriptscriptstyle QD}=
  {\cal L}^{1/8}_{\rm \scriptscriptstyle FP}
.
\end{equation}
This in turn implies that ${\cal L}_{\rm \scriptscriptstyle QD}$ can
be written in a manifest Lindblad form in terms of a single generator
for each Cartesian direction. We make the 
choice~\cite{art3,znf}
\begin{displaymath}
  {\hat {\bbox{a}}}_i=
{
\sqrt{2}
\over
 \lambda_M
}
\left(
{\hat {\bbox{x}}}_i
+{i\over\hbar}
{
\lambda_M^2
\over
                    4
}
{\hat {\bbox{p}}}_i
\right),
\end{displaymath}
where $
\lambda_{M}=\sqrt{\hbar^2\beta / M}
$,
the thermal wavelength associated to the Brownian particle, is put
into evidence, so that one has the commutation relations 
\begin{displaymath}
[
{\hat {\bbox{a}}}_i  ,
{\hat {\bbox{a}}}_j^{\scriptscriptstyle\dagger}
]
=\delta_{ij}
.  
\end{displaymath}
In such a way we have the alternative expression
\begin{equation}
  \label{last}
  {\cal L}_{\rm \scriptscriptstyle QD}[\cdot]=
          -
        {
        D_{pp}  
        \over
         \hbar^2
        }
        {
        \lambda_M^2
        \over
                 4
        }
        \sum_{i=1}^3
        \frac i\hbar
        \left[  
        \left \{  
         {\hat {\bbox{x}}}_i , 
        {\hat {\bbox{p}}}_i
        \right \}            
        ,\cdot
        \right]
        +
        {
        D_{pp}
        \over
         \hbar^2
        }
        \lambda_M^2
        \sum_{i=1}^3
        \left[  
        {
        {\hat {\bbox{a}}}_i
        \cdot
        {\hat {\bbox{a}}}_i^{\scriptscriptstyle\dagger}
        - { {1\over 2}}
        \{
        {\hat {\bbox{a}}}_i^{\scriptscriptstyle\dagger}
        {\hat {\bbox{a}}}_i
        ,\cdot
        \}
        }
        \right]      ,
\end{equation}
in which the single generator feature is put into evidence.
\section{CONCLUSIONS AND REMARKS}  
\label{cinque} 
In this paper we have considered the behavior with respect to
covariance under translations and rotations, and existence of a
stationary solution of a recently proposed master equation (\ref{me})
for the description of the interaction of a test particle with a
fluid, a physical model corresponding to the so-called Rayleigh
gas. The key result in  (\ref{me}) is the appearance of a two-point
correlation function known as dynamic structure factor and given by
(\ref{dsf}), the general structure conforming to results already
obtained in the mathematical literature for the generator of a
translation-covariant dynamical semigroup. This correlation function,
depending on symmetry and statistical mechanics properties of the
fluid, directly determines the behavior of the master equation with
respect to covariance under translations and rotations, and existence
of a stationary solution with the expected canonical form. Considering 
the specific case of a free gas, the dynamic structure factor has been
explicitly evaluated for Bose, Fermi and Maxwell-Boltzmann statistics, 
and the dependence on the physical parameters determining the
peculiar features of the model under consideration has been put into
evidence in an exact expansion of the dynamic structure
function. These parameters are the fugacity of the gas $z$, the ratio 
between mass of the gas particles and of the test particle $\alpha$,
the transferred momentum ${\bbox{q}}$ and the transferred energy $E
({\bbox{q}},{\bbox{p}})$. Stability of the covariance properties of
the master equation and of existence of a stationary solution is then
considered in the limit in which these parameters are small, together
with the different explicit expressions of the master equation. In
particular in the Brownian limit $\alpha\ll 1$ and considering small
momentum transfer, corresponding through the physical interpretation
of the dynamic structure factor to long-wavelength fluctuations, a
Fokker-Planck equation with a Lindblad structure is obtained, given by 
(\ref{compatto}), where the results corresponding to Bose, Fermi and
Maxwell-Boltzmann statistics are jointly considered. The correction
due to quantum statistics in the Fokker-Planck equation is simply
expressed through the Bose and Fermi functions given by (\ref{42}) and 
(\ref{44}) respectively.
\section*{ACKNOWLEDGMENTS}
The author would like to thank Prof. L. Lanz for 
his support during the whole work and Prof. A.
Barchielli  and Prof.~A.~S.~Holevo for useful suggestions. He also
thanks Dr.~F.~Belgiorno for careful reading of the manuscript.
This work was  supported by MURST under
Cofinanziamento and Progetto Giovani.
\appendix
\section{DERIVATION OF EQ.~(\ref{6}) AND EQ.~(\ref{7})}  
\label{aa} 
In this Appendix we want to explicitly calculate the expression of the 
dynamic structure factor for a free gas as a function of ${\bbox{q}}$
and ${\bbox{p}}$.
Working at finite temperature we can carry out the calculation for
both Bose and Fermi particles at the same time adopting the convention 
that the symbol $\pm$ means a $+$ sign for Bose particles and $-$ for
Fermi particles. We start from the expression
\begin{displaymath}
        S_{\rm \scriptscriptstyle B/F}({\bbox{q}},{\bbox{p}})
        =
        \frac 1n
        \int_{{\bf R}^3} 
        {
        d^3 \!
        {\bbox{k}}
        \over
        (2\pi \hbar)^3
        }
        \,  
        \langle
        n_{k}
        \rangle_{\rm \scriptscriptstyle B/F}
        (1\pm
        \langle
        n_{k - q}
        \rangle_{\rm \scriptscriptstyle B/F}
        )
        \delta       
        \left(
        {({{\bbox{p}} + {\bbox{q}}})^2\over 2M}
        +
        {({{\bbox{k}} - {\bbox{q}}})^2\over 2m}
        -
        {{{\bbox{p}}}^2 \over 2M}
        -
        {{\bbox{k}}^2\over 2m}
        \right)
\end{displaymath}
with
\begin{displaymath}
          \langle
        n_{k}
        \rangle_{\rm \scriptscriptstyle B/F}
        =
        {
        1
        \over  
        z^{-1}e^{\beta \epsilon_k}
        \mp 1}
      ,
      \qquad
      \epsilon_k = {{\bbox{k}}^2\over 2m}
      ,
\end{displaymath}
which can be found for example in~\cite{Lovesey} and corresponds to
(\ref{dsf}) for a free gas apart from a singular term proportional, in 
the continuum limit considered here, to
$\delta^3 ({\bbox{q}})$,
relevant only for $\bbox{q}=0$ and not contributing to the
master equation. In fact in the derivation of the master equation the
contributions for ${\bbox{q}}=0$ exactly cancel out. This term
according to (\ref{diff}) corresponds to 
forward scattering.
We now have to evaluate the integral in ${\bbox{k}}$.
This is most easily done writing $S_{\rm \scriptscriptstyle
  B/F}({\bbox{q}},{\bbox{p}})$ in the form
\begin{displaymath}
        S_{\rm \scriptscriptstyle B/F}({\bbox{q}},{\bbox{p}})
        =
        \frac 1n
        \int_{{\bf R}^3} 
        {
        d^3 \!
        {\bbox{k}}
        \over
        (2\pi \hbar)^3
        }
        \,  
        \langle
        n_{k}
        \rangle_{\rm \scriptscriptstyle B/F}
        (1\pm
        \langle
        n_{k - q}
        \rangle_{\rm \scriptscriptstyle B/F}
        )
        \delta       
        \left(
          E ({\bbox{q}},{\bbox{p}})
          + \epsilon_{k-q}-\epsilon_k
        \right)
\end{displaymath}
and observing that
\begin{displaymath}
        \langle
         n_{k}
        \rangle_{\rm \scriptscriptstyle B/F}
        (1\pm
        \langle
        n_{k - q}
        \rangle_{\rm \scriptscriptstyle B/F}
        )
        =
        {
        1  
        \over  
        1-e^{\beta (\epsilon_{k} - \epsilon_{k - q})}
        }
        (
        \langle
         n_{k}
        \rangle_{\rm \scriptscriptstyle B/F}
        -
        \langle
        n_{k - q}
        \rangle_{\rm \scriptscriptstyle B/F}
        )
        ,
\end{displaymath}
so that one has
\begin{eqnarray}
\label{12}
        S_{\rm \scriptscriptstyle B/F}({\bbox{q}},{\bbox{p}})
        &=&
        \frac 1n
        {
        1
        \over  
        1-e^{\beta E({\bbox{q}},{\bbox{p}})}
        }
        \int_{{\bf R}^3} 
        {
        d^3 \!
        {\bbox{k}}
        \over
        (2\pi \hbar)^3
        }
        \,  
        (
        \langle
         n_{k}
        \rangle_{\rm \scriptscriptstyle B/F}
        -
        \langle
        n_{k - q}
        \rangle_{\rm \scriptscriptstyle B/F}
        )
        \delta       
        \left(
          E ({\bbox{q}},{\bbox{p}})
          + \epsilon_{k-q}-\epsilon_k
        \right)
        \nonumber\\
        &=&
        \frac 1n
        {
        1
        \over  
        1-e^{\beta E({\bbox{q}},{\bbox{p}})}
        }
        \int_{{\bf R}^3} 
        {
        d^3 \!
        {\bbox{k}}
        \over
        (2\pi \hbar)^3
        }
        \, 
        \left\{
        \langle
         n_{k}
        \rangle_{\rm \scriptscriptstyle B/F}
        \delta       
        \left(
          E ({\bbox{q}},{\bbox{p}})
          + \epsilon_{k-q}-\epsilon_k
        \right)
        \right.
        \nonumber\\
        &&
        \hphantom{
        \frac 1n
        {
        1
        \over  
        1-e^{\beta E({\bbox{q}},{\bbox{p}})}
        }
        \int_{{\bf R}^3} 
        {
        d^3 \!
        {\bbox{k}}
        \over
        (2\pi \hbar)^3
        }
        \, 
        \langle
         n_{k}
        \rangle_{\rm \scriptscriptstyle B/F}
          }
        \left.
        -
        \langle
        n_{k}
        \rangle_{\rm \scriptscriptstyle B/F}
        \delta       
        \left(
          E ({\bbox{q}},{\bbox{p}})
          + \epsilon_{k}-\epsilon_{k+q}
        \right)
        \right\}
        \nonumber
        \\
        &=&
        \frac 1n
        {
        2m
        \over  
        1-e^{\beta E({\bbox{q}},{\bbox{p}})}
        }
        \int_{{\bf R}^3} 
        {
        d^3 \!
        {\bbox{k}}
        \over
        (2\pi \hbar)^3
        }
        \, 
         \langle
        n_{k}
        \rangle_{\rm \scriptscriptstyle B/F}
       \left\{
        \delta\left(
          2mE ({\bbox{q}},{\bbox{p}})
          + q^2-2{\bbox{k}}\cdot{\bbox{q}}
        \right)
        \right.
        \\
        &&
        \hphantom{
        \frac 1n
        {
        1
        \over  
        1-e^{\beta E({\bbox{q}},{\bbox{p}})}
        }
        \int_{{\bf R}^3} 
        {
        d^3 \!
        {\bbox{k}}
        \over
        (2\pi \hbar)^3
        }
        \, 
        \langle
         n_{k}
        \rangle_{\rm \scriptscriptstyle B/F}
          }
        \left.
        -
        \delta\left(
          2mE ({\bbox{q}},{\bbox{p}})
          -q^2-2{\bbox{k}}\cdot{\bbox{q}}
        \right)
        \right\}
.
        \nonumber
\end{eqnarray}
We are thus led to consider an integral of the form
\begin{equation}
  \label{10}
        \int_{{\bf R}^3} 
        {
        d^3 \!
        {\bbox{k}}
        }
        \, 
         \langle
        n_{k}
        \rangle_{\rm \scriptscriptstyle B/F}
        \delta
        \left(
        \eta-2{\bbox{k}}\cdot{\bbox{q}}
        \right)
\end{equation}
with $\eta$ a real parameter. Denoting by $\xi$ the cosine of the
angle between ${\bbox{k}}$ and ${\bbox{q}}$ the integral in (\ref{10}) 
becomes
\begin{eqnarray}
  \label{13}
        2\pi
        \int_{-1}^{1}
        d\xi
        \int_{0}^{+\infty}
        dk \, k^2
         \langle
        n_{k}
        \rangle_{\rm \scriptscriptstyle B/F}
        \delta
        \left(
        \eta-2\xi kq
        \right)
        &=&
        \int_{-1}^{1}
        d\xi
        \int_{0}^{+\infty}
        dk \, k^2
         \langle
        n_{k}
        \rangle_{\rm \scriptscriptstyle B/F}
        \int_{-\infty}^{+\infty}
        dp
        \,
        e^{ip (\eta-2\xi kq) }
        \nonumber
        \\
        &=&
        \int_{-\infty}^{+\infty}
        dk \, k
         \langle
        n_{k}
        \rangle_{\rm \scriptscriptstyle B/F}
        \int_{-\infty}^{+\infty}
        dp
        \,
        {
        e^{ip (\eta+2kq) }
        \over
        i2pq
        }
\end{eqnarray}
and using the identity
\begin{equation}
  \label{14}
  k         \langle
        n_{k}
        \rangle_{\rm \scriptscriptstyle B/F}
=
\pm
\frac{m}{\beta}
\frac{d}{dk}
\log [1\mp z e^{-\epsilon_k}]
\end{equation}
we get, integrating by parts,
\begin{eqnarray}
  \label{11}
        \int_{{\bf R}^3} 
        {
        d^3 \!
        {\bbox{k}}
        }
        \, 
         \langle
        n_{k}
        \rangle_{\rm \scriptscriptstyle B/F}
        \delta
        \left(
        \eta-2{\bbox{k}}\cdot{\bbox{q}}
        \right)
        &=&
        \mp
        \frac{2\pi m}{\beta}
        \int_{-\infty}^{+\infty}
        dk
        \,
        \log [1\mp z e^{-\epsilon_k}]
        \int_{-\infty}^{+\infty}
        \frac{dp}{2\pi}
        \,
        e^{ip (\eta+2kq) }
        \nonumber
        \\
        &=&
        \mp
        \frac{\pi m}{\beta q}
        \log \left\{1\mp z \exp{\left[ -\frac{\beta}{8m} 
            \left(\frac{\eta}{q}\right)^2\right]}\right\}
        .
\end{eqnarray}
Inserting the result (\ref{11}) in (\ref{12}) one immediately has
\begin{equation}
  \label{abf}
        S_{\rm \scriptscriptstyle B/F}({\bbox{q}},{\bbox{p}})
        =
        \mp
        {
        1
        \over
         (2\pi\hbar)^3
        }
        {
        2\pi m^2
        \over
        n\beta q
        }
        {
        1
        \over
        1-
        e^{\beta E({\bbox{q}},{\bbox{p}})}
        }
        \log
        \left[
        {
        1\mp z
        \exp{
        \left[
        -{
        \beta
        \over
             8m
        }
        {
        (2mE({\bbox{q}},{\bbox{p}}) + q^2)^2
        \over
                  q^2
        }
        \right]
        }
        \over
        1\mp z
        \exp{
        \left[
        -{
        \beta
        \over
             8m
        }
        {
        (2mE({\bbox{q}},{\bbox{p}}) - q^2)^2
        \over
                  q^2
        }
        \right]
        }
        }
        \right].
\end{equation}
In a similar way starting from the expression of a gas of
Maxwell-Boltzmann particles
\begin{displaymath}
        S_{\rm \scriptscriptstyle MB}({\bbox{q}},{\bbox{p}})
        =
        \frac 1n
        \int_{{\bf R}^3} 
        {
        d^3 \!
        {\bbox{k}}
        \over
        (2\pi \hbar)^3
        }
        \,  
        \langle
        n_{k}
        \rangle_{\rm \scriptscriptstyle MB}
        \delta       
        \left(
        {({{\bbox{p}} + {\bbox{q}}})^2\over 2M}
        +
        {({{\bbox{k}} - {\bbox{q}}})^2\over 2m}
        -
        {{{\bbox{p}}}^2 \over 2M}
        -
        {{\bbox{k}}^2\over 2m}
        \right)
        ,
\end{displaymath}
with
\begin{displaymath}
        \langle
        n_{k}
        \rangle_{\rm \scriptscriptstyle MB}
        =
        ze^{-\beta\epsilon_k},
\end{displaymath}
we write it in the form
\begin{eqnarray}
  \label{15}
        S_{\rm \scriptscriptstyle MB}({\bbox{q}},{\bbox{p}})
        &=&
        \frac 1n
        \int_{{\bf R}^3} 
        {
        d^3 \!
        {\bbox{k}}
        \over
        (2\pi \hbar)^3
        }
        \,  
        \langle
        n_{k}
        \rangle_{\rm \scriptscriptstyle MB}
        \delta       
        \left(
          E ({\bbox{q}},{\bbox{p}})
          + \epsilon_{k-q}-\epsilon_k
        \right)
        \nonumber
        \\
        &=&
        \frac{2m}{n}
        \int_{{\bf R}^3} 
        {
        d^3 \!
        {\bbox{k}}
        \over
        (2\pi \hbar)^3
        }
        \,  
        \langle
        n_{k}
        \rangle_{\rm \scriptscriptstyle MB}
        \delta\left(
          2mE ({\bbox{q}},{\bbox{p}})
          + q^2-2{\bbox{k}}\cdot{\bbox{q}}
        \right)
        .
\end{eqnarray}
Analogously to (\ref{10}) we have to consider 
\begin{displaymath}
        \int_{{\bf R}^3} 
        {
        d^3 \!
        {\bbox{k}}
        }
        \, 
         \langle
        n_{k}
        \rangle_{\rm \scriptscriptstyle MB}
        \delta
        \left(
        \eta-2{\bbox{k}}\cdot{\bbox{q}}
        \right),
\end{displaymath}
which according to (\ref{13}) becomes
\begin{displaymath}
          \int_{-\infty}^{+\infty}
        dk \, k
         \langle
        n_{k}
        \rangle_{\rm \scriptscriptstyle MB}
        \int_{-\infty}^{+\infty}
        dp
        \,
        {
        e^{ip (\eta+2kq) }
        \over
        i2pq
        }
      .
\end{displaymath}
Exploiting instead of (\ref{14}) the relation
\begin{displaymath}
   k         \langle
        n_{k}
        \rangle_{\rm \scriptscriptstyle MB}
=
-
\frac{m}{\beta}
\frac{d}{dk}
z e^{-\epsilon_k}
\end{displaymath}
we obtain
\begin{displaymath}
        \int_{{\bf R}^3} 
        {
        d^3 \!
        {\bbox{k}}
        }
        \, 
         \langle
        n_{k}
        \rangle_{\rm \scriptscriptstyle MB}
        \delta
        \left(
        \eta-2{\bbox{k}}\cdot{\bbox{q}}
        \right)
        =
        \frac{\pi m}{\beta q}z
        \exp{\left[ -\frac{\beta}{8m} \left(\frac{\eta}{q}\right)^2\right]}
        ,
\end{displaymath}
which has to be substituted in (\ref{15}) leading to
\begin{equation}
  \label{amb}
        S_{\rm \scriptscriptstyle MB}({\bbox{q}},{\bbox{p}})
        =
        {
        1
        \over
         (2\pi\hbar)^3
        }
        {
        2\pi m^2
        \over
        n\beta q
        }
        z
        \exp\left[
        -{
        \beta
        \over
             8m
        }
        {
        (2mE({\bbox{q}},{\bbox{p}}) + q^2)^2
        \over
                  q^2
        }
        \right]
        .
\end{equation}
\section{EXACT EXPANSION OF $S_{\rm \scriptscriptstyle
  B/F}$ AND DERIVATION OF EQ.~(\ref{40})}  
\label{bb}
We will now derive an expression for $S_{\rm \scriptscriptstyle
  B/F}({\bbox{q}},{\bbox{p}})$ equivalent to (\ref{8}), in which
however a series expansion in powers of the fugacity $z$ is put into
evidence. The starting point is the Taylor expansion for the logarithm
$\log (1+x)=\sum_{k=1}^{\infty} {(-)^{k+1} }  
x^k/k$, which leads to write
(\ref{8}) in the form
\begin{eqnarray}
  \label{b1}
        S_{\rm \scriptscriptstyle B/F}({\bbox{q}},{\bbox{p}})
        &=&
        \mp
        {
        1
        \over
         (2\pi\hbar)^3
        }
        {
        2\pi m^2
        \over
        n\beta q
        }
        {
        1
        \over
        1-
        e^{
          {
        \beta
        \over
             2m
        }
        \left(
        2\sigma({\bbox{q}},{\bbox{p}})q -q^2
        \right)
        }
      }
      \nonumber
      \\
      &&
      \hphantom{
        \mp
        {
        1
        \over
         (2\pi\hbar)^3
        }
        }
      \times
      \sum_{k=1}^{\infty} (-)^{k-1}{(\mp z)^k \over k}
      e^{
      -{
        \beta
        \over
             2m
        }
        \sigma^2({\bbox{q}},{\bbox{p}})
        }
        \left[
          1 -
        e^{
        k{
        \beta
        \over
             2m
        }
        (2\sigma({\bbox{q}},{\bbox{p}})q - q^2)
        }
      \right]
      .
\end{eqnarray}
Considering now in (\ref{b1}) a geometrical progression of reason
$        e^{
        {
        \beta
        \over
             2m
        }
        (2\sigma({\bbox{q}},{\bbox{p}})q - q^2)
        }
$, according to the formula
\begin{equation}
  \label{geo}
  1-x^k = (1-x) \sum_{n=0}^{k-1} x^n
\end{equation}
eq.~(\ref{b1}) becomes
\begin{displaymath}
        S_{\rm \scriptscriptstyle B/F}({\bbox{q}},{\bbox{p}})
        =
        {
        1
        \over
         (2\pi\hbar)^3
        }
        {
        2\pi m^2
        \over
        n\beta q
        }
      \sum_{k=1}^{\infty}(\pm)^{k+1} {z^k \over k}
      e^{  
      -k{
        \beta
        \over
             2m
        }
        \sigma^2({\bbox{q}},{\bbox{p}})
        }
      \sum_{n=0}^{k-1}      
        e^{
        n{
        \beta
        \over
             2m
        }
        (2\sigma({\bbox{q}},{\bbox{p}})q - q^2)
        }
      ,
\end{displaymath}
where it is to be noted that the sum over $n$ is due to the presence
of the statistical correction 
$        1\pm
        \langle
        n_{k - q}
        \rangle_{\rm \scriptscriptstyle B/F}
$ in (\ref{sbf}) and disappears, being replaced by a factor one, if
this correction is neglected. It is worthwhile to put into evidence a
factor
\begin{displaymath}
        {
        1
        \over
         (2\pi\hbar)^3
        }
        {
        2\pi m^2
        \over
        n\beta q
        }
      z
      e^{
        -{
        \beta
        \over
             2m
        }
        \sigma^2({\bbox{q}},{\bbox{p}})
        }
      ,
\end{displaymath}
corresponding to the expression of the dynamic structure factor for a
free gas of Maxwell-Boltzmann particles, thus obtaining
\begin{displaymath}
        S_{\rm \scriptscriptstyle B/F}({\bbox{q}},{\bbox{p}})
        =
        S_{\rm \scriptscriptstyle MB}({\bbox{q}},{\bbox{p}})
        \left[ {
          1 +
        \sum_{k=1}^{\infty}(\pm)^{k} {z^k \over k+1}
        e^{
        -k{
        \beta
        \over
             2m
        }
        \sigma^2({\bbox{q}},{\bbox{p}})
        }
      \sum_{n=0}^{k}      
        e^{
        n{
        \beta
        \over
             2m
        }
        (2\sigma({\bbox{q}},{\bbox{p}})q - q^2)
        }
      } \right]
      .
\end{displaymath}
In the Brownian limit $\alpha\ll 1$ considered in
Section~\ref{quattro}, neglecting in (\ref{20}) the contributions of
order $\alpha$, this
expression goes simply over to
\begin{equation}
  \label{b4}
        S^{\scriptscriptstyle\infty}_{\rm \scriptscriptstyle B/F}
        ({\bbox{q}},{\bbox{p}})
        =
        S^{\scriptscriptstyle\infty}_{\rm \scriptscriptstyle MB}
        ({\bbox{q}},{\bbox{p}})
        \left[ {
          1 +
        \sum_{k=1}^{\infty}(\pm)^{k} {z^k \over k+1}
        e^{        -k{
        \beta
        \over
             8m
        }
        q^2
        }
        e^{
        -k\frac{\beta}{2}
        [
        {
        q^2
        \over
             2M
        }
        +
        {{\bbox{q}}\cdot{\bbox{p}} \over M}        
        ]
        }
      \sum_{n=0}^{k}      
        e^{
        n\beta
        [
        {
        q^2
        \over
             2M
        }
        +
        {{\bbox{q}}\cdot{\bbox{p}} \over M}        
        ]
        }
      } \right]
      ,
\end{equation}
where 
$S^{\scriptscriptstyle\infty}_{\rm \scriptscriptstyle B/F}
({\bbox{q}},{\bbox{p}})$
and
$S^{\scriptscriptstyle\infty}_{\rm \scriptscriptstyle MB}
({\bbox{q}},{\bbox{p}})$ are given
respectively by (\ref{21}) and  (\ref{22}). In terms of $E
({\bbox{q}},{\bbox{p}})$ equation (\ref{b4}) takes the remarkably
compact form
\begin{equation}
  \label{b5}
        S^{\scriptscriptstyle\infty}_{\rm \scriptscriptstyle B/F}
        ({\bbox{q}},E)
        =
        S^{\scriptscriptstyle\infty}_{\rm \scriptscriptstyle MB}
        ({\bbox{q}},E)
        \left[ {
          1 +
        \sum_{k=1}^{\infty}(\pm)^{k} {z^k \over k+1}
        e^{        -k{
        \beta
        \over
             8m
        }
        q^2
        }
        e^{
        -k\frac{\beta}{2}E
        }
      \sum_{n=0}^{k}      
        e^{
        n\beta E
        }
      }\right]
      ,
\end{equation}
where 
$S^{\scriptscriptstyle\infty}_{\rm \scriptscriptstyle B/F}({\bbox{q}},E)$
and
$S^{\scriptscriptstyle\infty}_{\rm \scriptscriptstyle
  MB}({\bbox{q}},E)$ are given by (\ref{23}) and  (\ref{24}). 
We now go one step further noting that the following identity holds:
\begin{displaymath}
        e^{
        -k\frac{\beta}{2}E
        }
      \sum_{n=0}^{k}      
        e^{
        n\beta E
        }
      =
{
\sinh \left[(k+1)\frac{\beta}{2}E\right]
\over
\sinh \left(\frac{\beta}{2}E\right)
}
   ,
\end{displaymath}
which can be easily obtained exploiting (\ref{geo}), so that
(\ref{b5}) becomes
\begin{equation}
  \label{sviluppo}
        S^{\scriptscriptstyle\infty}_{\rm \scriptscriptstyle B/F}
        ({\bbox{q}},E)
        =
        S^{\scriptscriptstyle\infty}_{\rm \scriptscriptstyle MB}
        ({\bbox{q}},E)
        \sum_{k=0}^{\infty}(\pm)^{k} {z^k \over k+1}
        e^{        -k{
        \beta
        \over
             8m
        }
        q^2
        }
      {
        \sinh \left[(k+1)\frac{\beta}{2}E\right]
        \over
        \sinh \left(\frac{\beta}{2}E\right)
        }
      .
\end{equation}
Eq.~(\ref{sviluppo}) is the most convenient expression in order to
consider the limit of small momentum transfer. Exploiting the
expansion
\begin{displaymath}
      {
        \sinh \left[\frac{1}{2}(k+1)\beta E\right]
        \over
        \sinh \left(\frac{1}{2}\beta E\right)
        }
      =
      (k+1)\left[1+\frac{1}{24}\left( \beta E\right)^2
        (k^2+2k)+{\mathord{\mathrm{O}}} (E^4)\right] 
\end{displaymath}
and recalling that $E$ is given by (\ref{E}) we may write
(\ref{sviluppo})  as 
\begin{eqnarray}
  \label{b8}
        S^{\scriptscriptstyle\infty}_{\rm \scriptscriptstyle B/F}
        ({\bbox{q}},E)
        &=&
        S^{\scriptscriptstyle\infty}_{\rm \scriptscriptstyle MB}
        ({\bbox{q}},E)
        \left[
          \sum_{k=0}^{\infty} (\pm z)^k -
          \frac{\beta}{8m} q^2\sum_{k=1}^{\infty}(\pm)^k k z^k
          \right.
          \\
          &&
          \hphantom{{S^{\scriptscriptstyle\infty}_{\rm \scriptscriptstyle MB}
        ({\bbox{q}},E)
          \sum_{k=0}^{\infty} 
          }}
        \left.
          + \frac{1}{12} (\beta E)^2\sum_{k=1}^{\infty}(\pm)^k k z^k
          + \frac{1}{24} (\beta E)^2\sum_{k=1}^{\infty}(\pm)^k k^2 z^k
          + {\mathord{\mathrm{O}}} (q^4)
        \right]
        \nonumber
        .
\end{eqnarray}
Recalling (\ref{sviluppo}) and the explicit expression of 
$S^{\scriptscriptstyle\infty}_{\rm \scriptscriptstyle MB}$ given by
(\ref{sbf}) one also has
\begin{displaymath}
        S^{\scriptscriptstyle\infty}_{\rm \scriptscriptstyle B/F}
        ({\bbox{q}},E)
        =
        \pm
        {
        1
        \over
         (2\pi\hbar)^3
        }
        {
        2\pi m^2
        \over
        n\beta q
        }
        {
        e^{-\frac{\beta}{2}E}
        \over
        \sinh \left(\frac{\beta}{2}E\right)
        }
        \sum_{k=1}^{\infty}(\pm)^{k} {z^k \over k}
        e^{        -k{
        \beta
        \over
             8m
        }
        q^2
        }
        \sinh \left( k\frac{\beta}{2}E\right)
\end{displaymath}
and exploiting~\cite{Ryzhik}
\begin{displaymath}
        \sum_{k=1}^{\infty}{p^k \over k}
        \sinh (kx)
        =
        \mathop{\mathrm{arth}}\nolimits
        \left[
        \frac{p\sinh x}{1-p\cosh x}
        \right]  
\end{displaymath}
we obtain the alternative expression
\begin{equation}
\label{nuova}
        S^{\scriptscriptstyle\infty}_{\rm \scriptscriptstyle B/F}
        ({\bbox{q}},E)
        =
        \pm
        {
        1
        \over
         (2\pi\hbar)^3
        }
        {
        2\pi m^2
        \over
        n\beta q
        }
        {
        e^{-\frac{\beta}{2}E}
        \over
        \sinh \left(\frac{\beta}{2}E\right)
        }
        \mathop{\mathrm{arth}}\nolimits
        \left[
        {
        \pm z         
        e^{        -{
        \beta
        \over
             8m
        }
        q^2
        }
       \sinh \left( \frac{\beta}{2}E\right)
         \over
         1        \mp z         e^{        -{
        \beta
        \over
             8m
        }
        q^2
        }
       \cosh \left( \frac{\beta}{2}E\right)
         }
        \right]  
\end{equation}
equivalent to (\ref{21}) as can also be directly checked starting from
the identity
\begin{equation}
  \label{arco}
        \mathop{\mathrm{arth}}\nolimits x = \frac{1}{2} \log
        \left[ \frac{1+x}{1-x} \right]
        .
\end{equation}
Note that  (\ref{21}) and (\ref{nuova}) in the Fermi case can also be
written in the form
\begin{displaymath}
        S^{\scriptscriptstyle\infty}_{\rm \scriptscriptstyle F}
        ({\bbox{q}},E)
        =
        -
        {
        1
        \over
         (2\pi\hbar)^3
        }
        {
        \pi m^2
        \over
        n\beta q
        }
        {
        e^{-\frac{\beta}{2}E}
        \over
        \sinh \left(\frac{\beta}{2}E\right)
        }
         \log
        \left[
        {
        1- \frac{z}{1+z}
        \left(
        1 -  
        e^{
        -{
        \beta
        \over
             8m
        }
        q^2
        }
        e^{
        -\frac{\beta}{2}E
        }
        \right)
        \over
        1- \frac{z}{1+z}
        \left(
         1 - 
        e^{
        -{
        \beta
        \over
             8m
        }
        q^2
        }
        e^{
        +\frac{\beta}{2}E
        }
        \right)
        }
        \right]
\end{displaymath}
and
\begin{displaymath}
        S^{\scriptscriptstyle\infty}_{\rm \scriptscriptstyle F}
        ({\bbox{q}},E)
        =
        -
        {
        1
        \over
         (2\pi\hbar)^3
        }
        {
        2\pi m^2
        \over
        n\beta q
        }
        {
        e^{-\frac{\beta}{2}E}
        \over
        \sinh \left(\frac{\beta}{2}E\right)
        }
      \mathop{\mathrm{arth}}\nolimits 
        \left[
          {
          \frac{z}{1+z}
        e^{
        -{
        \beta
        \over
             8m
        }
        q^2
        }
      \sinh \left(\frac{\beta}{2}E\right)
        \over
        \frac{z}{1+z}
        \left(
        1 -
        e^{
        -{
        \beta
        \over
             8m
        }
        q^2
        }
      \cosh \left(\frac{\beta}{2}E\right)
      \right) 
      -1
         }
        \right]
\end{displaymath}
respectively, that can be useful if one is interested in an expansion
for large values of $z$.
According to (\ref{arco}) also for $S_{\rm
  \scriptscriptstyle B/F}({\bbox{q}},E)$ given by (\ref{6}) one has
the alternative expression 
\begin{equation}
\label{arth}
        S_{\rm \scriptscriptstyle B/F}
        ({\bbox{q}},E)
        =
        \pm
        {
        1
        \over
         (2\pi\hbar)^3
        }
        {
        2\pi m^2
        \over
        n\beta q
        }
        {
        e^{-\frac{\beta}{2}E}
        \over
        \sinh \left(\frac{\beta}{2}E\right)
        }
        \mathop{\mathrm{arth}}\nolimits
        \left[
        {
        \pm z         
        e^{        -{
        \beta
        \over
             8m
        }
        q^2
        }
         e^{-\frac{\beta}{2}\frac{m}{q^2}E^2}
       \sinh \left( \frac{\beta}{2}E\right)
         \over
         1        \mp z         e^{        -{
        \beta
        \over
             8m
        }
        q^2
        }
        e^{-\frac{\beta}{2}\frac{m}{q^2}E^2}
       \cosh \left( \frac{\beta}{2}E\right)
         }
        \right]  
        .
\end{equation}
The validity of the detailed balance condition for (\ref{arth})
according to   (\ref{dbce}) can immediately be checked observing that
both $\sinh x$ and  $\mathop{\mathrm{arth}}\nolimits x$ are odd
functions, while $\cosh x$ is an even function.
  
\end{document}